\documentclass[graybox]{svmult}

\usepackage{natbib}

\usepackage{mathptmx}       
\usepackage{helvet}         
\usepackage{courier}        
\usepackage{type1cm}        
\usepackage{amssymb}                            
\usepackage{gensymb}
\usepackage{makeidx}         
\usepackage{graphicx}        
\usepackage{multicol}        
\usepackage[bottom]{footmisc}

\makeindex             

\newcommand{\msol}{\mbox{$M_\odot$}}

\newcommand{\cmjj}{\mbox{${\rm cm^{-2}}$}}
\newcommand{\HI}{\mbox{H\,{\scriptsize I}}}
\newcommand{\lya}{\mbox{${\rm Ly}\alpha$}}

\newcommand{\MgII}{{\mbox{Mg\,{\scriptsize II}}}}

\newcommand{\OVI}{{\mbox{O\,{\scriptsize VI}}}}
\newcommand{\CII}{{\mbox{C\,{\scriptsize II}}}}
\newcommand{\CIII}{{\mbox{C\,{\scriptsize III}}}}
\newcommand{\CIV}{{\mbox{C\,{\scriptsize IV}}}}
\newcommand{\SiII}{{\mbox{Si\,{\scriptsize II}}}}
\newcommand{\SiIII}{{\mbox{Si\,{\scriptsize III}}}}
\newcommand{\SiIV}{{\mbox{Si\,{\scriptsize IV}}}}

\begin{document}

\title*{Outskirts of Distant Galaxies In Absorption}
\author{Hsiao-Wen Chen}
\institute{Hsiao-Wen Chen \at Department of Astronomy \& Astrophysics, and Kavli Institute for
Cosmological Physics, The University of Chicago, Chicago, IL 60637, USA,  \email{hchen@oddjob.uchicago.edu}}
%
%
\maketitle

\abstract{QSO absorption spectroscopy provides a sensitive probe of
  both the neutral medium and diffuse ionized gas in the distant
  Universe.  It extends 21\,cm maps of gaseous structures around
  low-redshift galaxies both to lower gas column densities and to
  higher redshifts.  Combining galaxy surveys with absorption-line
  observations of gas around galaxies enables comprehensive studies of
  baryon cycles in galaxy outskirts over cosmic time.  This Chapter
  presents a review of the empirical understanding of the cosmic
  neutral gas reservoir from studies of damped \lya\ absorbers (DLAs).
  It describes the constraints on the star formation relation and
  chemical enrichment history in the outskirts of distant galaxies
  from DLA studies.  A brief discussion of available constraints on
  the ionized circumgalactic gas from studies of lower column density
  \lya\ absorbers and associated ionic absorption transitions is
  presented at the end.}

\section{Introduction}
\label{sec:intro}

\begin{figure}
\begin{center}
\includegraphics[scale=0.53]{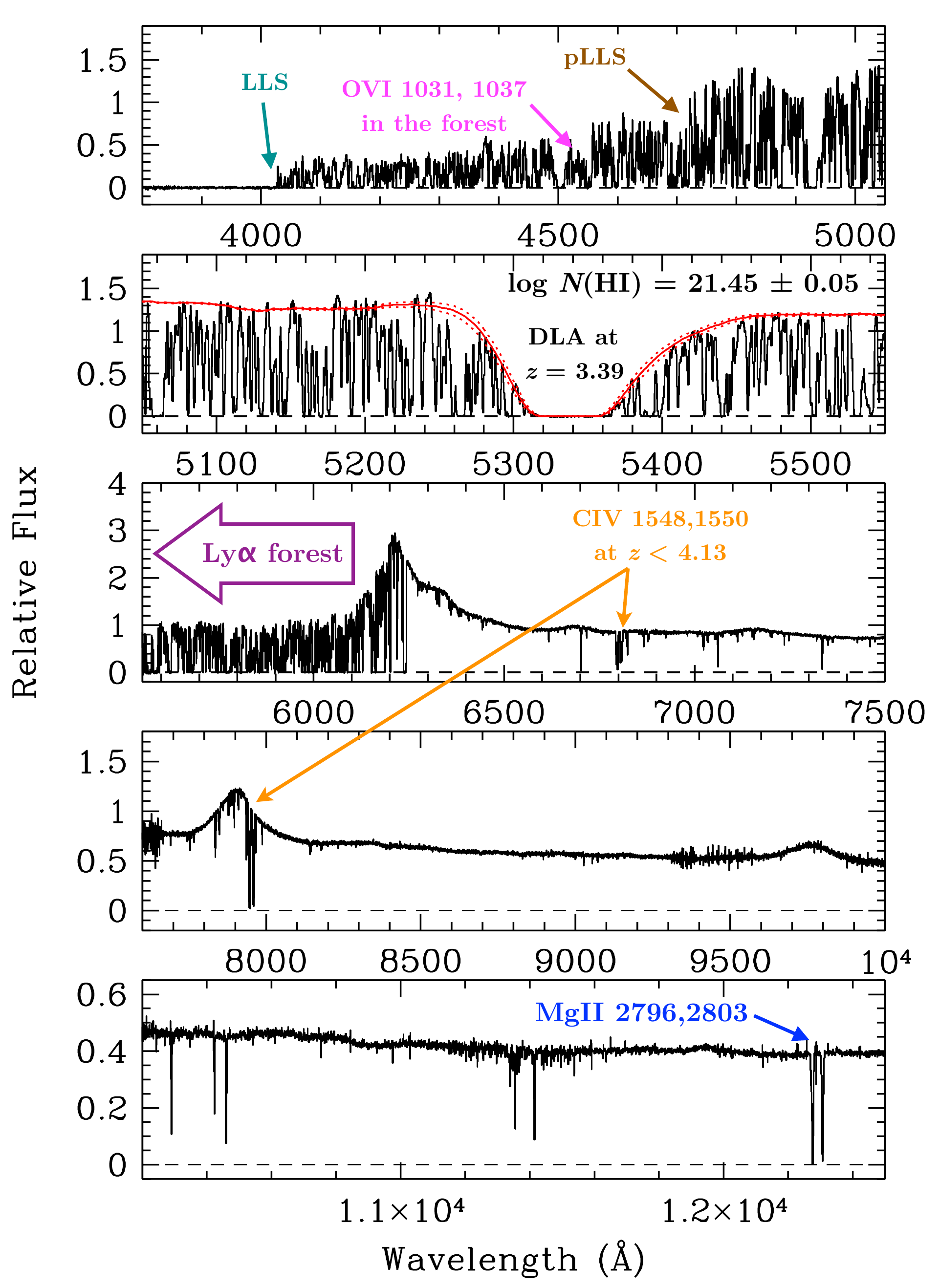}
\end{center}
\caption{Example of the wealth of information for intervening gas
  revealed in the optical and near-infrared spectrum of a QSO at
  $z=4.13$.  In addition to broad emission lines intrinsic to the QSO,
  such as \lya/N\,V at $\approx 6200$\,\AA, a forest of
  \lya\,$\lambda$\,1215 absorption lines\index{Lyman $\alpha$ forest} is
  observed blueward of 6200\,\AA.  These \lya\ forest lines arise in
  relatively high gas density regions at $z_{\rm abs}\lesssim  z_{\rm QSO
  }$ along the line of sight.  The \lya\ absorbers span over 10
  decades in neutral hydrogen column densities ($N({\HI})$), and
  include (1) neutral damped \lya\ absorbers (DLAs)\index{damped Lyman $\alpha$ absorbers}, 
  (2) optically thick Lyman limit systems
  (LLS)\index{Lyman limit systems}, (3) partial LLS (pLLS), and (4)
  highly ionized \lya\ absorbers (see text for a quantitative
  definition of these different classes).  The DLAs are characterized
  by pronounced damping wings ({\it second panel from the top}), while LLS
  and pLLS are identified based on the apparent flux discontinuities
  in QSO spectra ({\it top panel}).  Many of these strong \lya\ absorbers
  are accompanied with metal absorption transitions such as the
  \OVI\,$\lambda\lambda$\,1031, 1037 doublet transitions which occur
  in the \lya\ forest, and the \CIV\,$\lambda\lambda$\,1548, 1550 and
  \MgII\,$\lambda\lambda$\,2796, 2803 doublets.  Together, these metal
  lines constrain the ionization state and chemical enrichment of the
  gas}
\label{fig:qsospec}       
\end{figure}

Absorption-line spectroscopy\index{absorption-line spectroscopy}
complements emission surveys and provides a powerful tool for studying
the diffuse, large-scale baryonic structures in the distant Universe
(e.g., \citealt{Rauch:1998,Wolfe:2005,Prochaska:2009ASSP}).  Depending
on the physical conditions of the gas (including gas density,
temperature, ionization state, and metallicity), a high-density region
in the foreground is expected to imprint various absorption
transitions of different line strengths in the spectrum of a
background QSO.  Observing the absorption features imprinted in QSO
spectra enables a uniform survey of diffuse gas in and around
galaxies, as well as detailed studies of the physical conditions of
the gas at redshifts as high as the background sources can be
observed.

\begin{figure}
\begin{center}
\includegraphics[scale=0.33]{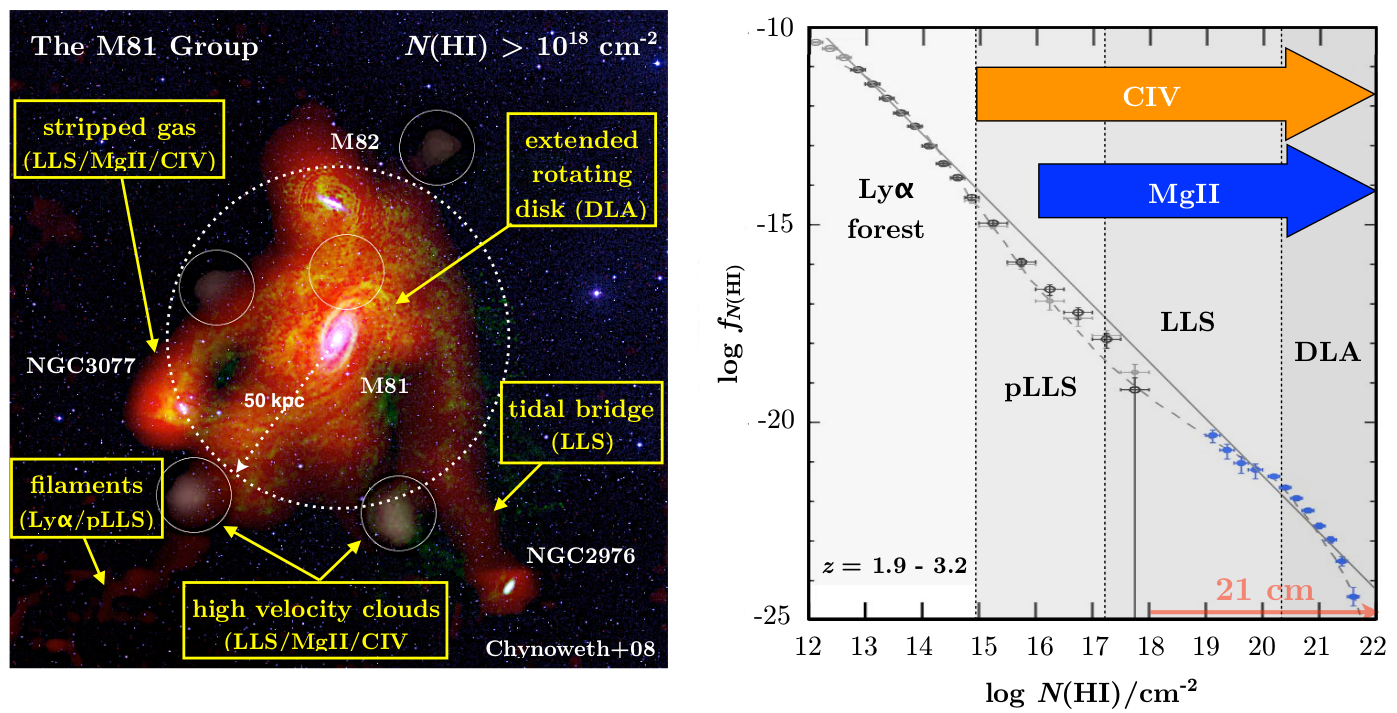}
\end{center}
\caption{Mapping galaxy outskirts in 21\,cm and in QSO absorption-line
  systems.  {\it Left}: Deep 21\,cm image of the M81 group, revealing a
  complex interface between stars and gas in the group.  The observed
  neutral hydrogen column densities range from $N({\HI})\sim
  10^{18}\,\cmjj$ in the filamentary structures to $N({\HI})>
  10^{21}\,\cmjj$ in the star-forming disks of group members
  (\citealt{Yun:1994, Chynoweth:2008}).  The 21\,cm image reveals a
  diverse array of gaseous structures in this galaxy group, but these
  observations become extremely challenging beyond redshift $z \approx
  0.2$ (e.g., \citealt{Verheijen:2007, Fernandez:2013}).  {\it Right}:
  The \HI\ column density distribution function\index{H\,{\sc i} column density distribution function} of \lya\ absorbers, $f_{N(\HI)}$,
  uncovered at $z=1.9-3.2$ along sightlines toward random background
  QSOs (adapted from \citealt{Kim:2013}).  Quasar absorbers in different
  categories are mapped onto different \HI\ structures both seen and
  missed in the 21\,cm image in the left panel.  Specifically, DLAs
  probe the star-forming ISM and extended rotating disks, LLS probe
  the gaseous streams connecting different group members as well as
  stripped gas and high velocity clouds around galaxies, and pLLS and
  strong \lya\ absorbers trace ionized gas that is not observed in
  21\,cm signals.  Among the quasar absorbers, \CIV\ absorption
  transitions are commonly observed in strong \lya\ absorbers of
  $N({\HI})\gtrsim  10^{15}\,\cmjj$ (see e.g., \citealt{Kim:2013,
    Dodorico:2016}), and \MgII\ absorption transitions are seen in
  most high-$N({\HI})$ absorbers of $N({\HI})\gtrsim  10^{16}\,\cmjj$ (see
  e.g., \citealt{Rigby:2002}).  These metal-line absorbers trace
  chemically enriched gas in and around galaxies}
\label{fig:halomap}       
\end{figure}

Figure~\ref{fig:qsospec} displays an example of optical and
near-infrared spectra of a high-redshift QSO.  The QSO is at redshift
$z_{\rm QSO}=4.13$, and the spectra are retrieved from the XQ-100
archive (\citealt{Lopez:2016}).  At the QSO redshift, multiple broad
emission lines are observed, including the \lya/N\,V emission at
$\approx 6200$\,\AA, \CIV\ emission at $\approx 7900$\,\AA, and \CIII]
  emission at $\approx 9800$\,\AA.  Blueward of the \lya\ emission line
  are a forest of \lya\,$\lambda$\,1215 absorption lines\index{Lyman $\alpha$ forest}
  produced by intervening overdense regions at $z_{\rm
    abs}\lesssim  z_{\rm QSO }$ along the QSO sightline.  These overdense
  regions span a wide range in \HI\ column density ($N({\HI})$), from
  neutral interstellar gas of $N({\HI})\ge 10^{20.3}\,\cmjj$, to
  optically opaque Lyman limit systems (LLS) of
  $N({\HI})>10^{17.2}\,\cmjj$, to optically thin partial LLS (pLLS)
  with $N({\HI})=10^{15-17.2}\,\cmjj$, and to highly ionized
  \lya\ forest lines with $N({\HI})=10^{12-15}\,\cmjj$ (right panel of
  Fig.~\ref{fig:halomap}).

The large $N({\HI})$ in the neutral medium produces pronounced damping
wings in the QSO spectrum.  These absorbers are commonly referred to
as damped \lya\ absorbers (DLAs)\index{damped Lyman $\alpha$ absorbers}.  An
example is shown in the second panel from the top in Fig.~\ref{fig:qsospec}.  In this particular case, a simultaneous fit to the
QSO continuum and the damping wings (red curve in the second panel
from the top) yields a best-fit $\log\,N({\rm \HI})=21.45\pm 0.05$
for the DLA.  At intermediate $N(\HI)$, LLS and pLLS are identified
based on the apparent flux discontinuities in QSO spectra (top panel).
A significant fraction of these strong \lya\ absorbers have been
enriched with heavy elements which produce additional absorption
features due to heavy ions in the QSO spectra.  The most prominent
features include the \OVI\,$\lambda\lambda$\,1031, 1037 doublet
transitions which occur in the \lya\ forest, and the
\CIV\,$\lambda\lambda$\,1548, 1550 and \MgII\,$\lambda\lambda$\,2796,
2803 doublets, plus a series of low-ionization transitions such as
C\,{\scriptsize II}, Si\,{\scriptsize II}, and Fe\,{\scriptsize II}.
Together, these ionic transitions constrain the ionization state and
chemical compositions of the gas (e.g., \citealt{Chen:2000, Werk:2014}).

Combining galaxy surveys with absorption-line observations of gas
around galaxies enables comprehensive studies of baryon cycles between
star-forming regions and low-density gas over cosmic time.  At low
redshifts, $z\lesssim  0.2$, deep 21\,cm and CO surveys have revealed
exquisite details of the cold gas content ($T\lesssim  1000$\,K) in nearby
galaxies, providing both new clues and puzzles in the overall
understanding of galaxy formation and evolution.  These include
extended \HI\ disks around blue star-forming galaxies with the
\HI\ extent $\approx 2\times$ what is found for the stellar disk
(e.g., \citealt{Swaters:2002, Walter:2008, Leroy:2008}), extended
\HI\ and molecular gas in early-type galaxies (e.g.,
\citealt{Oosterloo:2010, Serra:2012}) with predominantly old stellar
populations and little or no on-going star formation (\citealt{Salim:2010}),
and widespread \HI\ streams connecting regular-looking galaxies in
group environments (e.g., \citealt{Verdes:2001, Chynoweth:2008}).

Figure~\ref{fig:halomap} (left panel) showcases an example of a deep
21\,cm image of the M81 group, a poor group of dynamical mass $M_{\rm
  dyn}\sim 10^{12}\,M_\odot$ (\citealt{Karachentsev:2006}).
Prominent group members include the grand-design spiral galaxy M81 at
the centre, the proto-starburst galaxy M82, and several other
lower-mass satellites (\citealt{Burbidge:1961}).  The 21\,cm image
displays a diverse array of gaseous structures in the M81 group, from
extended rotating disks, warps, high velocity clouds (HVCs)\index{high velocity clouds}, 
tidal tails and filaments, to bridges connecting
what appear to be optically isolated galaxies.  High column density
gaseous streams of $N(\HI)\gtrsim  10^{18}\,\cmjj$ are seen extending
beyond 50\,kpc in projected distance from M81, despite the isolated
appearances of M81 and other group members in optical images.  These
spatially resolved imaging observations of different gaseous
components serve as important tests for theoretical models of galaxy
formation and evolution (e.g., \citealt{Agertz:2009, Marasco:2016}).
However, 21\,cm imaging observations are insensitive to warm ionized
gas of $T\sim 10^4$\,K and become extremely challenging for galaxies
beyond redshift $z=0.2$ (e.g., \citealt{Verheijen:2007, Fernandez:2013}).

QSO absorption spectroscopy extends 21\,cm maps of gaseous structures
around galaxies to both lower gas column density and higher redshifts.
Based on the characteristic $N(\HI)$, direct analogues can be drawn
between different types of QSO absorbers and different gaseous
components seen in deep 21\,cm images of nearby galaxies.  For example,
DLAs probe the neutral gas in the interstellar medium (ISM) and
extended rotating disks, LLS probe optically thick gaseous streams and
high velocity clouds in galaxy haloes, and pLLS and strong
\lya\ absorbers of $N({\HI})\approx 10^{14-17}\,\cmjj$ trace ionized
halo gas and starburst outflows\index{starburst outflows} (e.g.,
supergalactic winds in M\,82 \citealt{Lehnert:1999}) that cannot be
reached with 21\,cm observations.

The right panel of Fig.~\ref{fig:halomap} displays the \HI\ column
density distribution function, $f_{N(\HI)}$, for all \lya\ absorbers
uncovered at $z=1.9-3.2$ along random QSO sightlines
(\citealt{Kim:2013}).  $f_{N(\HI)}$, defined as the number of
\lya\ absorbers per unit absorption pathlength per unit \HI\ column
density interval, is a key statistical measure of the \lya\ absorber
population.  It represents a cross-section weighted surface density
profile of hydrogen gas in a cosmological volume.  With sufficiently
high spectral resolution and high signal-to-noise, $S/N\gtrsim  30$, QSO
absorption spectra probe tenuous gas with $N(\HI)$ as low as
$N(\HI)\sim 10^{12}\,\cmjj$.  The steeply declining $f_{N(\HI)}$ with
increasing $N(\HI)$ shows that the occurrence (or areal coverage) of
pLLS and strong \lya\ absorbers of $N({\HI})\approx 10^{14-17}\,\cmjj$
is $\approx 10$ times higher than that of optically thick LLS along a
random sightline and $\approx 100$ times higher than the incidence of
DLAs.  Such a differential frequency distribution is qualitatively
consistent with the spatial distribution of \HI\ gas recorded in local
21\,cm surveys (e.g., Fig.~\ref{fig:halomap}, left panel), where
gaseous disks with $N(\HI)$ comparable to DLAs cover a much smaller
area on the sky than streams and HVCs with $N(\HI)$ comparable to LLS.
If a substantial fraction of optically thin absorbers originate in
galaxy haloes, then their higher incidence implies a gaseous halo of
size at least three times what is seen in deep 21\,cm images.

In addition, many of these strong \lya\ absorbers exhibit associated
transitions due to heavy ions.  In particular, \CIV\ absorption
transitions are commonly observed in strong \lya\ absorbers of
$N({\HI})\gtrsim  10^{15}\,\cmjj$ (see, e.g., \citealt{Kim:2013,
  Dodorico:2016}), and \MgII\ absorption transitions are seen in most
high-$N({\HI})$ absorbers of $N({\HI})\gtrsim  10^{16}$ $\cmjj$ (e.g.,
\citealt{Rigby:2002}).  While \MgII\ absorbers\index{Mg\,{\sc ii} absorbers}
are understood to originate in photo-ionized gas of temperature $T\sim
10^4$\,K (e.g., \citealt{Bergeron:1986}),
\CIV\ absorbers\index{C\,{\sc iv} absorbers} are more commonly seen in
complex, multi-phase media (e.g., \citealt{Rauch:1996, Boksenberg:2015}).
These metal-line absorbers therefore offer additional probes of
chemically enriched gas in and around galaxies.

This Chapter presents a brief review of the current state of knowledge
on the outskirts of distant galaxies from absorption-line studies.
The review will first focus on the properties of the neutral gas
reservoir probed by DLAs, and then outline the insights into star
formation and chemical enrichment in the outskirts of distant galaxies
from searches of DLA galaxies.  A comprehensive review of DLAs is
already available in \cite{Wolfe:2005}.  Therefore, the emphasis here
focusses on new findings over the past decade.  Finally, a brief
discussion will be presented on the empirical properties and physical
understandings of the ionized circumgalactic gas as probed by strong
\lya\ and various metal-line absorbers.

\section{Tracking the Neutral Gas Reservoir Over Cosmic Time}
\label{sec:DLAstats}

DLAs\index{damped Lyman $\alpha$ absorbers} are historically defined as
\lya\ absorbers with neutral hydrogen column densities exceeding
$N(\HI)=2\times 10^{20}\,\cmjj$ (\citealt{Wolfe:2005}), corresponding to
a surface mass density limit of $\Sigma_{\rm atomic}\approx
2\,\msol\,{\rm pc}^{-2}$ for atomic gas (including helium).  The large
gas surface mass densities revealed in high-redshift DLAs are
comparable to what is seen in 21\,cm observations of nearby
star-forming galaxies (e.g., \citealt{Walter:2008, Leroy:2008}), making
DLAs a promising signpost of young galaxies in the distant Universe
(\citealt{Wolfe:1986}).  In addition, the $N(\HI)$ threshold ensures that
the gas is neutral under the metagalactic ionizing radiation field
(e.g., \citealt{Viegas:1995, Prochaska:1996, Prochaska:2002}).  Neutral
gas provides the seeds necessary for sustaining star formation.
Therefore, observations of DLAs not only help establish a census of
the cosmic evolution of the neutral gas reservoir (e.g.,
\citealt{Neeleman:2016}), but also offer a unique window into star
formation physics in distant galaxies (e.g., \citealt{Lanzetta:2002,
  Wolfe:2006}).

While the utility of DLAs for probing the young Universe is clear,
these objects are relatively rare (see the right panel of Fig.~\ref{fig:halomap}) and establishing a statistically representative
sample of these rare systems requires a large sample of QSO spectra.
Over the last decade, significant progress has been made in
characterizing the DLA population at $z\gtrsim  2$, owing to the rapidly
growing spectroscopic sample of high-redshift QSOs from the Sloan
Digital Sky Survey (SDSS; \citealt{York:2000}).  The blue points in the
right panel of Fig.~\ref{fig:halomap} are based on $\sim 1000$ DLAs
and $\sim 500$ strong LLS identified at $z\approx 2-5$ in an initial
SDSS DLA sample (\citealt{Noterdaeme:2009}).  The sample of known DLAs at
$z\gtrsim  2$ has continued to grow, reaching $\sim 10,000$ DLAs found in
the SDSS spectroscopic QSO sample (e.g., \citealt{Noterdaeme:2012}).

\begin{figure}
\begin{center}
\includegraphics[scale=0.33]{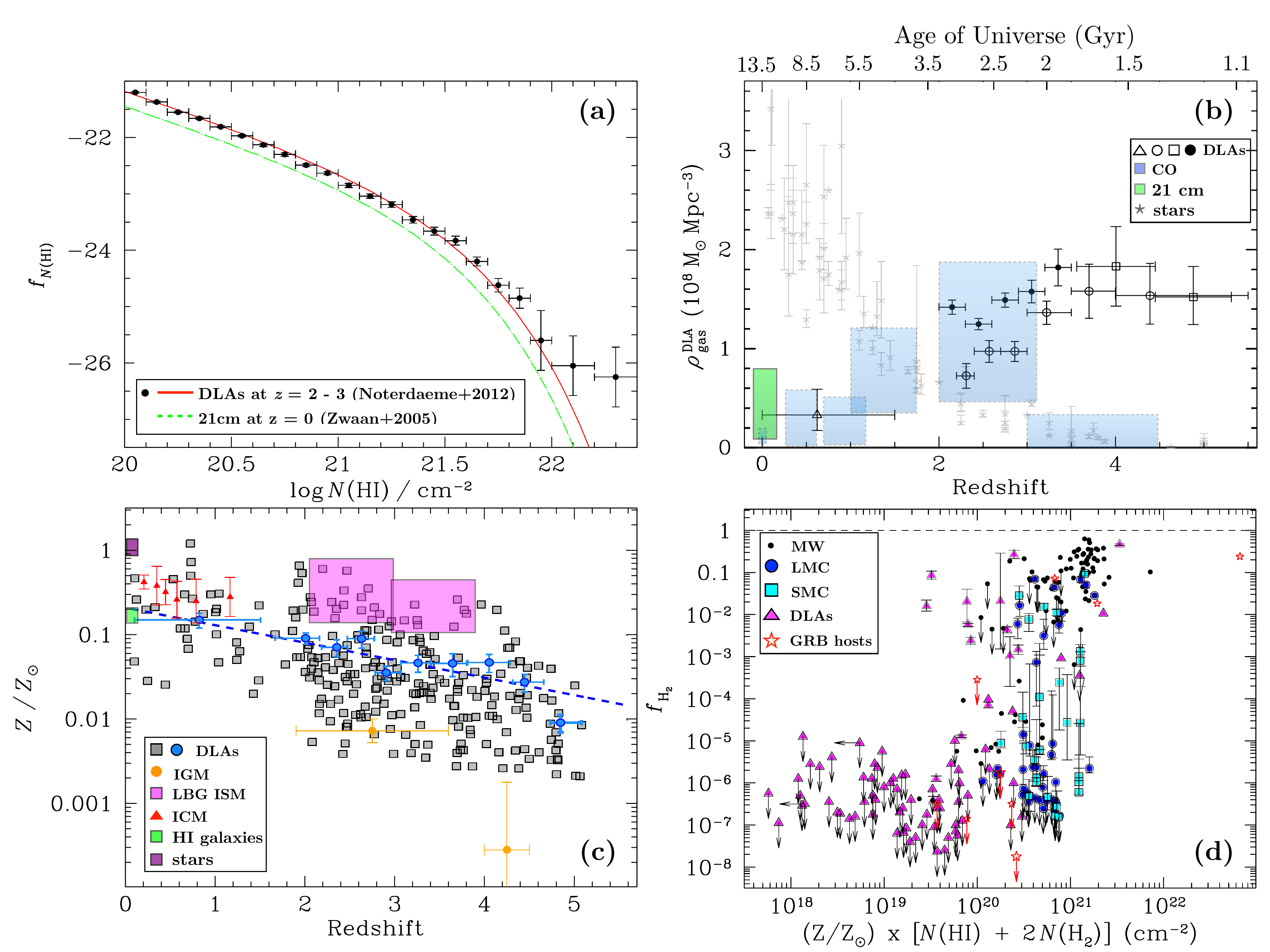}
\end{center}
\caption{Summary of known DLA properties: ({\it a}) evolving neutral
  hydrogen column density distribution functions\index{H\,{\sc i} column density distribution function}, $f_{N(\HI)}$ from DLAs at $z=2-3$
  (\citealt{Noterdaeme:2012}) to \HI\ galaxies at $z\approx 0$
  (\citealt{Zwaan:2005}); ({\it b}) declining cosmic neutral gas mass
  density\index{neutral gas mass density} with
  increasing Universe age (or decreasing redshift) from observations
  of DLAs (solid points from \citealt{Noterdaeme:2012}, open circles from
  \citealt{Prochaska:2009}, open squares from \citealt{Crighton:2015}, and
  open triangle from \citealt{Neeleman:2016}) following Eq.~(2), local \HI\ galaxies (green
  shaded box, a compilation from \citealt{Neeleman:2016}), and molecular
  gas (blue shaded boxes, \citealt{Decarli:2016}), in comparison to
  increasing cosmic stellar mass density in galaxies with increasing
  Universe age (grey asterisks, a compilation from \citealt{Madau:2014});
  ({\it c}) gas-phase metallicity\index{gas metallicity} ($Z$) relative to
  Solar ($Z_\odot$) as a function of redshift in DLAs (grey squares
  for individual absorbers and blue points for $N(\HI)$-weighted mean
  from Marc Rafelski, \citealt{Rafelski:2012, Rafelski:2014}), IGM at
  $z\gtrsim  2$ (orange circles, \citealt{Aguirre:2008, Simcoe:2011}), ISM of
  starburst galaxies at $z\approx 2-4$ (light magenta boxes,
  \citealt{Pettini:2001, Pettini:2004, Erb:2006, Maiolino:2008,
    Mannucci:2009}), intracluster medium in X-ray luminous galaxy
  clusters at $z\lesssim  1$ (red triangles, \citealt{Balestra:2007}),
  \HI-selected galaxies (green box, \citealt{Zwaan:2005}), and stars at
  $z=0$ (dark purple box, \citealt{Gallazzi:2008}); and ({\it d}) molecular
  gas fraction, $f_{\rm H_2}$\index{molecular gas fraction} versus total surface density of neutral
  gas scaled by gas metallicity for high-redshift DLAs in triangles
  (\citealt{Noterdaeme:2008, Noterdaeme:2016}), $\gamma$-ray burst host
  ISM in star symbols (e.g., \citealt{Noterdaeme:2015}), and local ISM in the
  Milky Way (\citealt{Wolfire:2008}) and Large and Small Magellanic
  Clouds (\citealt{Tumlinson:2002}) in dots, blue circles, and cyan
  squares, respectively}
\label{fig:dla}       
\end{figure}

The large number of known DLAs has led to an accurate characterization
of the neutral gas reservoir at high redshifts.  Figure~\ref{fig:dla}a
displays the observed $N(\HI)$ distribution function\index{H\,{\sc i} column density distribution function}, 
$f_{\rm DLA}$, based on $\sim 7000$
DLAs identified at $z\approx 2-5$ (\citealt{Noterdaeme:2012}).  The plot
shows that $f_{\rm DLA}$ is well represented by a Schechter function
(\citealt{Schechter:1976}) at $\log\,N(\HI)\lesssim  22$ following
\begin{equation}
f_{\rm DLA}\equiv f_{N(\HI)}(\log\,N(\HI)\ge 20.3)\propto\left[\frac{N(\HI)}{N_*(\HI)}\right]^{\alpha}\,\exp[-N(\HI)/N_*(\HI)], 
\label{eq:fn}
\end{equation}
with a shallow power-law index of $\alpha\approx -1.3$ below the
characteristic \HI\ column density $\log\,N_*(\HI)\approx 21.3$ and a
steep exponential decline at larger $N(\HI)$ (\citealt{Noterdaeme:2009,
  Noterdaeme:2012}).  At $\log\,N(\HI)>22$, the observations clearly
deviate from the best-fit Schechter function.  However, DLAs are also
exceedingly rare in this high-$N(\HI)$ regime.  Only eight such strong
DLAs have been found in this large DLA sample
(\citealt{Noterdaeme:2012}), making measurements of $f_{\rm DLA}$ in the
two highest-$N(\HI)$ bins very uncertain.
In comparison to $f_{N(\HI)}$ established from 21\,cm maps of nearby
galaxies (\citealt{Zwaan:2005}), the amplitude of $f_{\rm DLA}$ at $z\gtrsim 
2$ is $\approx 2\times$ higher than $f_{N(\HI)}$ at $z\approx 0$ but
the overall shapes are remarkably similar at both low- and
high-$N(\HI)$ regimes (Fig.~\ref{fig:dla}a; see also
\citealt{Sanchez:2016, Rafelski:2016}).

At $\log\,N(\HI)>21$, numerical simulations have shown that the
predicted shape in $f_{\rm DLA}$ is sensitive to the detailed ISM
physics, including the formation of molecules ($H_2$) and different
feedback processes (e.g., \citealt{Altay:2011, Altay:2013, Bird:2014}).
Comparison of the observed and predicted $f_{\rm DLA}$ therefore
provides an independent and critical test for the prescriptions of
these physical processes in cosmological simulations.  However, the
constant exponentially declining trend at $N(\HI)\gtrsim  2\times
10^{21}\,\cmjj$ between low-redshift \HI\ galaxies and high-redshift
DLAs presents a puzzle.

At $z=0$, the rapidly declining $f_{N(\HI)}$ at $N(\HI)\gtrsim  N_*(\HI)$
has been interpreted as due to the conversion of atomic gas to
molecular gas (\citealt{Zwaan:2006, Braun:2012}).  As illustrated at the
end of this Section and in Fig.~\ref{fig:dla}d, the column density
threshold beyond which the gas transitions from \HI\ to ${\rm H_2}$
depends strongly on the gas metallicity, and the mean metallicity
observed in the atomic gas decreases steadily from $z\approx 0$ to
$z>4$ (Fig.~\ref{fig:dla}c).  Therefore, the conversion to molecules
in high-redshift DLAs is expected to occur at higher $N(\HI)$,
resulting in a higher $N_*(\HI)$ with increasing redshift.  However,
this is not observed (e.g., \citealt{Prochaska:2009, Sanchez:2016,
  Rafelski:2016}; Fig.~\ref{fig:dla}a).  Based on spatially resolved
21\,cm maps of nearby galaxies with ISM metallicity spanning over a
decade, it has been shown that $f_{N(\HI)}$ established individually
for these galaxies does not vary significantly with their ISM
metallicity (\citealt{Erkal:2012}).  Together, these findings demonstrate
that the exponential decline of $f_{\rm DLA}$ at $N(\HI)\gtrsim  N_*(\HI)$
is not due to conversion of \HI\ to ${\rm H_2}$, but the physical
origin remains unknown.

Nevertheless, the observed $f_{\rm DLA}$ immediately leads to two
important statistical quantities: (1) the number density of DLAs per
unit survey pathlength, obtained by integrating $f_{\rm DLA}$ over all
$N(\HI)$ greater than $N_0= 2\times 10^{20}\,\cmjj$ and (2) the cosmic
neutral gas mass density\index{neutral gas mass density}, contained in DLAs, $\Omega_{\rm atomic}$, which is the
$N(\HI)$-weighted integral of $f_{\rm DLA}$ following 
\begin{equation}
\label{omega}
\Omega_{\rm
  atomic}\equiv\rho_{\rm gas}/\rho_{\rm
  crit}=\int_{N_0}^\infty\,(\mu\,H_0/c/\rho_{\rm
  crit})\,N(\HI)\,f_{\rm DLA}\,d\,N(\HI),
\end{equation}
 where $\mu=1.3$ is the mean
atomic weight of the gas particles (accounting for the presence of
helium), $H_0$ is the Hubble constant, $c$ is the speed of light, and
$\rho_{\rm crit}$ is the critical density of the Universe (e.g.,
\citealt{Lanzetta:1991, Wolfe:1995}).  The shallow power-law index
$\alpha$ in the best-fit $f_{\rm DLA}$, together with a steep
exponential decline at high $N(\HI)$ from the Schechter function in
Eq.~(\ref{eq:fn}), indicates that while DLAs of $N(\HI)<N_*(\HI)$
dominate the neutral gas cross-section (and therefore the number
density), strong DLAs of $N(\HI)\sim N_*(\HI)$ contribute
predominantly to the neutral mass density in the Universe (e.g.,
\citealt{Zwaan:2005}).  A detailed examination of the differential
$\Omega_{\rm atomic}$ distribution as a function of $N(\HI)$ indeed
confirms that the bulk of neutral gas is contained in DLAs of
$N(\HI)\approx 2\times 10^{21}\,\cmjj$ (e.g., \citealt{Noterdaeme:2012}).

The cosmic evolution of $\rho_{\rm gas}$ observed in DLAs, from Eq.~(2), is shown in
black points in Fig.~\ref{fig:dla}b.  Only measurements based on
blind DLA surveys are presented in the plot\footnote{At $z\lesssim  1.6$,
  DLA surveys require QSO spectroscopy carried out in space and have
  been limited to the number of UV-bright QSOs available for
  absorption line searches.  Consequently, the number of known DLAs
  from blind surveys is small, $\approx 15$ (see \citealt{Neeleman:2016}
  for a compilation).  To increase substantially the sample of known
  DLAs at low redshifts, Rao \& Turnshek (\citealt{Rao:2006}) devised a
  clever space programme to search for new DLAs in known
  \MgII\ absorbers.  Their strategy yielded a substantial gain,
  tripling the total sample size of $z\lesssim  1.6$ DLAs.  However, the
  \MgII-selected DLA sample also includes a survey bias that is not
  well understood.  It has been shown that excluding \MgII-selected
  DLAs reduces the inferred $\Omega_{\rm atomic}$ by more than a
  factor of four (e.g., \citealt{Neeleman:2016}).  For consistency, only
  measurements of $\Omega_{\rm atomic}$ based on blind DLA surveys are
  included in the plot.}.  These include an early sample of $\approx
700$ DLAs at $z=2.5-5$ in the SDSS Data Release (DR) 5 (open circles;
\citealt{Prochaska:2009}), an expanded sample of $\approx 7000$ DLAs in
the SDSS DR12 (solid points; \citealt{Noterdaeme:2012}), an expanded
high-redshift sample of DLAs at $z=4-5$ (open squares;
\citealt{Crighton:2015}), and a sample of $\approx 14$ DLAs at $z\lesssim 
1.6$ from an exhaustive search in the {\it Hubble Space Telescope}
({\it HST}) UV spectroscopic archive (open triangle;
\citealt{Neeleman:2016}).

A range of mean \HI\ mass density at $z\approx 0$ has been reported
from different 21\,cm surveys (see \citealt{Neeleman:2016} for a recent
compilation).  These measurements are included in the green box in
Fig.~\ref{fig:dla}b.  Despite a relatively large scatter between
different 21\,cm surveys and between DLA surveys, a steady decline in
$\Omega_{\rm atomic}$ is observed from $z\approx 4$ to $z\approx 0$.
For comparison, the cosmic evolution of the molecular gas mass
density\index{molecular gas mass density} obtained
from a recent blind CO survey (\citealt{Decarli:2016}) is also included
as blue-shaded boxes in Fig.~\ref{fig:dla}b, along with the cosmic
evolution of stellar mass density measured in different galaxy surveys,
shown in grey asterisks (data from \citealt{Madau:2014}).  Figure~\ref{fig:dla}b shows that the decline in the neutral gas mass density
with decreasing redshift is coupled with an increase in the mean
stellar mass density in galaxies, which is qualitatively consistent
with the expectation that neutral gas is being consumed to form stars.
However, it is also clear that atomic gas alone is insufficient to
explain the observed order-of-magnitude gain in the total stellar mass
density from $z\approx 3$ to $z\approx 0$, which implies the need for
replenishing the neutral gas reservoir with accretion from the intergalactic medium (IGM)
(e.g., \citealt{Keres:2009, Prochaska:2009}).  At the same time, new
blind CO surveys have shown that molecular gas contributes roughly an
equal amount of neutral gas mass density as atomic gas observed in
DLAs at $z\lesssim  3$ (e.g., \citealt{Walter:2014, Decarli:2016}), although
the uncertainties are still very large.  Together with the knowledge
of an extremely low molecular gas fraction in DLAs (see the discussion
on the next page and Fig.~\ref{fig:dla}d), these new CO surveys
indicate that previous estimates of the total neutral gas mass density
based on DLAs alone have been underestimated by as much as a factor of
two.  An expanded blind CO survey over a cosmological volume is needed
to reduce the uncertainties in the observed molecular gas mass
densities at different redshifts, which will cast new insights into
the connections between star formation, the neutral gas reservoir, and the
ionized IGM over cosmic time.

Observations of the chemical compositions of DLAs provide additional
clues to the connection between the neutral gas probed by DLAs and
star formation (e.g., \citealt{Pettini:2004}).  In particular, because
the gas is predominantly neutral, the dominant ionization for most
heavy elements (such as Mg, Si, S, Fe, Zn, etc.) are in the
singly ionized state and therefore the observed abundances of these
low-ionization species place direct and accurate constraints on the
elemental abundances of the gas (e.g., \citealt{Viegas:1995,
  Prochaska:1996, Vladilo:2001, Prochaska:2002}).  Additional
constraints on the dust content and on the sources that drive the
chemical enrichment history\index{chemical enrichment history} in DLAs
can be obtained by comparing the relative abundances of different
elements.  Specifically, comparing the relative abundances between
refractory (such as Cr and Fe) and non-refractory elements (such as S
and Zn) indicates the presence of dust in the neutral gas, the amount
of which increases with metallicity (e.g., \citealt{Meyer:1989,
  Pettini:1990, Savage:1996, Wolfe:2005}).  The relative abundances of
$\alpha$- to Fe-peak elements determine whether core-collapse
supernovae (SNe) or SNe~Ia dominate the chemical enrichment history,
and DLAs typically exhibit an $\alpha$-element enhanced abundance
pattern (e.g., \citealt{Lu:1996, Pettini:1999, Prochaska:1999}).

Figure~\ref{fig:dla}c presents a summary of gas metallity\index{gas metallicity} ($Z$) relative to Solar ($Z_\odot$) measured for $>250$
DLAs at $z\lesssim  5$ (grey squares from \citealt{Rafelski:2012,
  Rafelski:2014}).  The cosmic mean gas metallicity in DLAs as a
function of redshift can be determined based on a $N(\HI)$-weighted
average over an ensemble of DLAs in each redshift bin (blue points),
which is found to increase steadily with decreasing redshift following
a best-fit mean relation of $\langle\,Z/Z_\odot\,\rangle=[-0.20\pm
  0.03]\,z-[0.68\pm 0.09]$ (dashed blue line, \citealt{Rafelski:2014}).
For comparison, the figure also includes measurements for stars (dark
purple box, \citealt{Gallazzi:2008}) and \HI-selected galaxies (green
box, \citealt{Zwaan:2005}) at $z=0$, iron abundances in the intracluster
medium in X-ray luminous galaxy clusters at $z\lesssim  1$ (red triangles,
\citealt{Balestra:2007}), ISM of starburst galaxies (light magenta boxes)
at $z\approx 2-3$ (\citealt{Pettini:2001, Pettini:2004, Erb:2006}) and at
$z=3-4$ (\citealt{Maiolino:2008, Mannucci:2009}), and IGM at $z\gtrsim  2$
(orange circles, \citealt{Aguirre:2008, Simcoe:2011}).

It is immediately clear from Fig.~\ref{fig:dla}c that there exists a
large scatter in the observed metallicity in DLAs at all redshifts.
In addition, while the cosmic mean metallicity in DLAs is
significantly higher than what is observed in the low-density IGM, it
remains lower than what is observed in the star-forming ISM at $z=2-4$ and 
a factor of $\approx 5$ below the mean values observed in
stars at $z=0$.  The chemical enrichment level in DLAs is also lower
than the iron abundances seen in the intracluster medium at
intermediate redshifts.  The observed low metallicity relative to the
measurements in and around known luminous galaxies raised the question
of whether or not the DLAs probe preferentially low-metallicity,
gas-rich galaxies and are not representative of more luminous,
metal-rich galaxies found in large-scale surveys (e.g.,
\citealt{Pettini:2004}).

The large scatter in the observed metallicity in DLAs is found to be
explained by a combination of two factors (\citealt{Chen:2005}): (i) the
mass-metallicity\index{mass-metallicity relation} (or
luminosity-metallicity) relation in which more massive galaxies on
average exhibit higher global ISM metallicities (e.g.,
\citealt{Tremonti:2004, Erb:2006, Neeleman:2013, Christensen:2014}) and
(ii) metallicity gradients\index{metallicity gradient} commonly seen
in star-forming disks with lower metallicities at larger distances
(e.g., \citealt{Zaritsky:1994, vanZee:1998, Sanchez:2014, Wuyts:2016}).
If DLAs sample a representative galaxy population including both
low-mass and massive galaxies and probe both inner and outer disks of
these galaxies, then a large metallicity spread is expected.

The observed low metallicity in DLAs, relative to star-forming ISM, is
also understood as due to a combination of DLAs being a gas
cross-section selected sample and the presence of metallicity
gradients in disk galaxies (\citealt{Chen:2005}).  A cross-section
selected sample contains a higher fraction of absorbers originating in
galaxy outskirts than in the inner regions, and the presence of
metallicity gradients indicates that galaxy outskirts have lower
metallicities than what is observed in inner disks (see Sect.~\ref{sec:DLAgals} and Fig.~\ref{fig:outskirts} below for more
details).  Indeed, including both factors, a gas cross-section
weighting scheme and a metallicity gradient, for local \HI\ galaxies
resulted in a mean metallicity comparable to what is observed in DLAs
(green box in Fig.~\ref{fig:dla}c; \citealt{Zwaan:2005}).

While DLAs exhibit a moderate level of chemical enrichment, searches
for molecular gas in DLAs have yielded only a few detections (e.g.,
\citealt{Noterdaeme:2008, Jorgenson:2014, Noterdaeme:2016}).  Figure~\ref{fig:dla}d displays the observed molecular gas fraction, which is
defined as $f_{\rm H_2}\equiv 2\,N({\rm H_2})/[N(\HI)+2\,N({\rm
    H_2})]$, versus metallicity-scaled total hydrogen column density
for $\approx 100$ DLAs at $z\approx 2-4$ (triangles).  The DLAs span
roughly two decades in $N(\HI)$ from $N(\HI)\approx 2\times
10^{20}\,\cmjj$ to $N(\HI)\approx 2.5\times 10^{22}\,\cmjj$.  Strong
limits have been placed for $f_{\rm H_2}$ for the majority of DLAs at
$f_{\rm H_2}\lesssim  10^{-5}$ with only $\approx 10$\% displaying the
presence of ${\rm H}_2$\index{molecular hydrogen} and two having $f_{\rm
  H_2}>0.1$.  In contrast, the ISM of the Milky Way (MW), at
comparable $N(\HI)$, displays a much higher $f_{\rm H_2}$ than the
DLAs at high redshifts.

The formation of molecules is understood to depend on two competing
factors: (i) the ISM radiation field which photo-dissociates molecules
and (ii) dust which facilitates molecule formation (e.g.,
\citealt{Elmegreen:1993, Cazaux:2004}).  Dust is considered a more
dominant factor because of its dual roles in both forming molecules
and shielding them from the ISM radiation field.  In star-forming
galaxies, the dust-to-gas mass ratio is observed to correlate strongly
with ISM gas-phase metallicity (e.g., \citealt{Leroy:2011, Remy:2014}).
It is therefore expected that the observed molecular gas fraction
should correlate with gas metallicity (e.g., \citealt{Elmegreen:1989,
  Krumholz:2009, Gnedin:2009}).

In the MW ISM with metallicity roughly Solar, $Z\approx Z_\odot$, the
molecular gas fraction is observed to increase sharply from $f_{\rm
  H_2}<10^{-4}$ to $f_{\rm H_2}\gtrsim  0.1$ at $N(\HI)\approx 2\times
10^{20}\,\cmjj$ (see \citealt{Wolfire:2008}).  The sharp transition from
atomic to molecular is also observed in the ISM of the Large and Small
Magellanic Clouds (LMC and SMC), but occurs at higher gas column
densities of $N(\HI)\approx 10^{21}\,\cmjj$ for the LMC and
$N(\HI)\approx 3\times 10^{21}\,\cmjj$ for the SMC (see
\citealt{Tumlinson:2002}).  The ISM metallicities of LMC and SMC are
$Z\approx 0.5\,Z_\odot$ and $Z\approx 0.15\,Z_\odot$, respectively.
These observations therefore support a simple metallicity-dependent
transitional gas column density illustrated in Fig.~\ref{fig:dla}d.
Following the metallicity-scaling relation, it is clear that despite a
high $N(\HI)$, most DLAs do not have sufficiently high metallicity
(and therefore dust content) to facilitate the formation of molecules
(\citealt{Gnedin:2010, Gnedin:2014, Noterdaeme:2015}).  This finding also
applies to $\gamma$-ray burst (GRB) host galaxies (star symbols in
Fig.~\ref{fig:dla}d).  With few exceptions (\citealt{Prochaska:2009GRB,
  Kruhler:2013, Friis:2015}, the ISM in most GRB hosts displays a combination
of very high $N(\HI)$ and low $f_{\rm H_2}$ (e.g.,
\citealt{Tumlinson:2007, Ledoux:2009}).  The observed absence of ${\rm
  H}_2$ in DLAs, together with a large molecular mass density revealed
in blind CO surveys (e.g., \citealt{Walter:2014, Decarli:2016}), shows
that a complete census for the cosmic evolution of the neutral gas
reservoir requires complementary surveys of molecular gas over a broad
redshift range.  In addition, as described in Sect.~\ref{sec:sfr}
below, the observed low molecular gas content also has important
implications for star formation properties in metal-deficient, high
neutral gas surface density environments.

\section{Probing the Neutral Gas Phase in Galaxy Outskirts}
\label{sec:DLAgals}

Considerable details have been learned about the physical properties
and chemical enrichment in neutral atomic gas from DLA studies.  To
apply the knowledge of DLAs for a better understanding of distant
galaxies, it is necessary to first identify DLA galaxies and compare
them with the general galaxy population.  Searches for DLA galaxies
are challenging, because distant galaxies are faint and because the
relatively small extent of high-$N(\HI)$ gas around galaxies places
the absorbing galaxies at small angular distances from the bright
background QSOs.  Based on a well-defined \HI\ size-mass relation
observed in local \HI\ galaxies (e.g., \citealt{Broeils:1997,
  Verheijen:2001, Swaters:2002}), the characteristic projected
separation (accounting for weighting by cross section) between a DLA
and an $L_*$ absorbing galaxy is $\approx 16$\,kpc and smaller for
lower-mass galaxies.  At $z=1-2$, a projected distance of 16\,kpc
corresponds to an angular separation of $\lesssim  2''$, and greater at
lower and higher redshifts.

\begin{figure}
\begin{center}
\includegraphics[scale=0.33]{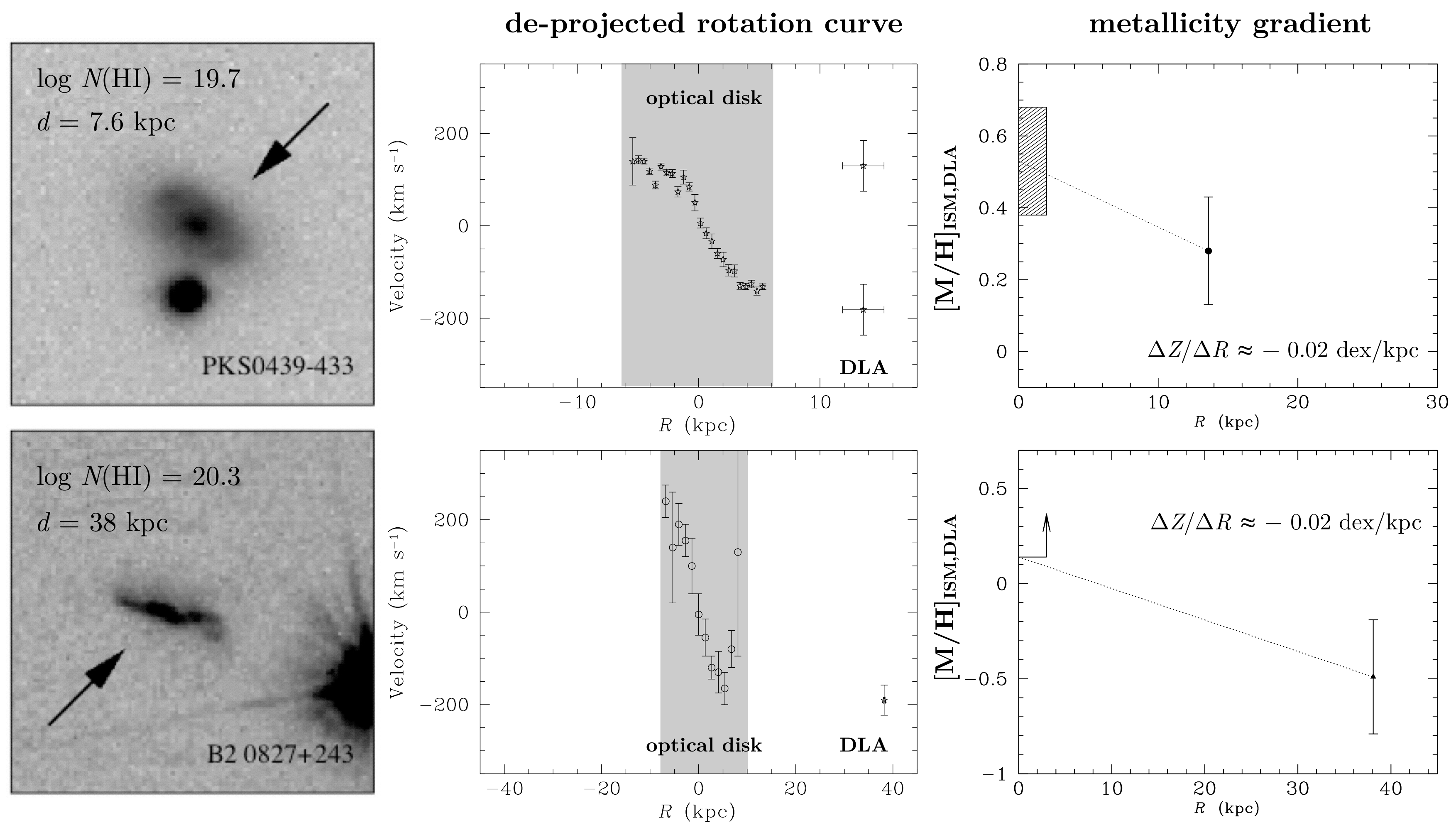}
\end{center}
\caption{Neutral gas kinematics and metallicity revealed by the
  presence of a DLA in the outskirts of two $L_*$ galaxies (adapted
  from \citealt{Chen:2005}).  The {\it top} row presents a DLA found at $d=7.6$\,kpc from a disk galaxy at $z=0.101$, which also exhibits widespread
  CO emission in the disk (\citealt{Neeleman:2016CO}).  The {\it bottom} row
  presents a DLA at $d=38$\,kpc from an edge-on disk at $z=0.525$.
  Deep $r$-band images of the galaxies are presented in the {\it left}
  panels, which display spatially resolved disk morphologies and
  enable accurate measurements of the inclination and orientation of
  the optical disk.  The {\it middle} panels present the optical rotation
  curves\index{rotation curve} deprojected along the disk plane
  (points in shaded area) based on the inclination angle determined
  from the optical image of each galaxy (Eq.~3 \& 4).  If the
  DLAs occur in extended disks, the corresponding galactocentric
  distances of the two galaxies from Eq.~(3) are $R=13.6$\,kpc (top)
  and $R=38$\,kpc (bottom).  The DLA in the {\it top} panel is resolved into
  two components of comparable ionic column densities
  (\citealt{Som:2015}) but an order of magnitude difference in $N({\rm
    H_2})$ (\citealt{Muzahid:2015}).  The component with a lower $N({\rm
    H_2})$ appears to be co-rotating with the optical disk (lower DLA
  data point), while the component with stronger $N({\rm H_2})$
  appears to be counter-rotating, possibly due to a satellite (upper
  DLA data point).  The DLA in the {\it bottom} panel displays simpler gas
  kinematics consistent with an extended rotating disk out to $\approx
  40$\,kpc.  The {\it right} panels present the metallicity
  gradient\index{metallicity gradient} observed in the gaseous disks
  based on comparisons of ISM gas-phase metallicity and metallicity of
  the DLA beyond the optical disks.  In both cases, the gas
  metallicity declines with increasing radius according to
  $\Delta\,Z/\Delta\,R=-0.02$ dex\,kpc$^{-1}$}
\label{fig:outskirts}       
\end{figure}

While fewer DLAs are known at $z\lesssim  1$ (see Sect.\ref{sec:DLAstats}), a large number ($\approx 40$) of these
low-redshift DLAs have their galaxy counterparts (or candidates) found
based on a combination of photometric and spectroscopic techniques
(e.g., \citealt{Chen:2003, Rao:2003, Rao:2011, Peroux:2016}).  It has
been shown based on this low-redshift DLA galaxy sample that DLAs
probe a representative galaxy population in luminosity and colour.  DLA
galaxies are consistent with an \HI\ cross-section selected sample
with a large fraction of DLAs found at projected distance $d\gtrsim  10$
kpc from the absorbing galaxies (e.g., \citealt{Chen:2003, Rao:2011}).
In addition to regular disk galaxies, two DLAs have been found in
a group environment (e.g., \citealt{Bergeron:1991, Chen:2003,
  Kacprzak:2010, Peroux:2011}), suggesting that stripped gas from
galaxy interactions could also contribute to the incidence of DLAs.
The low-redshift DLA sample is expected to continue to grow
dramatically with new discoveries from the SDSS (e.g.,
\citealt{Straka:2015}).  In contrast, the search for DLA galaxies at
$z>2$ has been less successful despite extensive efforts (e.g.,
\citealt{Warren:2001, Moller:2002, Peroux:2012, Fumagalli:2015}).  To
date, only $\approx 12$ DLA galaxies have been found at $z>2$ (\citealt{Krogager:2012, Fumagalli:2015}).

In addition to a general characterization of the DLA galaxy
population, individual DLA and galaxy pairs provide a unique
opportunity to probe neutral gas in the outskirts of distant galaxies.
Figure~\ref{fig:outskirts} shows two examples of constraining the
kinematics and chemical enrichment in the outskirts of neutral disks
from combining resolved optical imaging and spectroscopy of the galaxy
with an absorption-line analysis of the DLA.  In the first example
(top row), a DLA of $\log\,N(\HI)=19.7$ is found at $d=7.6$\,kpc from
an $L_*$ galaxy at $z=0.101$, which also exhibits widespread CO
emission in the disk (\citealt{Neeleman:2016CO}).  The galaxy disk is
resolved in the ground-based $r$-band image (upper-left panel), which
enables accurate measurements of the disk inclination and orientation
(\citealt{Chen:2005}).  While the observed $N(\HI)$ falls below the
nominal threshold of a DLA, the gas is found to be largely neutral
(e.g., \citealt{Chen:2005, Som:2015}).  In addition, abundant ${\rm H_2}$
is detected in the absorbing gas (\citealt{Muzahid:2015}).  Optical
spectra of the galaxy clearly indicate a strong velocity shear along
the disk, suggesting an organized rotation motion (\citealt{Chen:2005})
which is confirmed by recent CO observations (\citealt{Neeleman:2016CO}).
At the same time, the DLA is resolved into two components of
comparable ionic column densities (\citealt{Som:2015}) but an order of
magnitude difference in $N({\rm H_2})$ (\citealt{Muzahid:2015}).  A
rotation curve of the gaseous disk extending beyond 10\,kpc (top-centre
panel) can be established based on the observed velocity shear
($v_{\rm obs}$) and deprojection onto the disk plane following
\begin{equation}
\frac{R}{d}=\sqrt{1+\sin^2(\phi)\tan^2(i)}
\end{equation}
and 
\begin{equation}
v=\frac{v_{\rm obs}}{\cos(\phi)\sin(i)}\sqrt{1+\sin^2(\phi)\tan^2(i)},
\end{equation}
where $R$ is the galactocentric radius
along the disk, $v$ is the deprojected rotation
velocity, $i$ is the inclination
angle of the disk, and $\phi$ is the azimuthal angle from the major
axis of the disk where the DLA is detected (\citealt{Chen:2005}, see also
\citealt{Steidel:2002} for an alternative formalism).  For the two
absorbing components in this DLA, it is found that the component with
a lower $N({\rm H_2})$ appears to be co-rotating with the optical disk
(lower DLA data point), while the component with stronger $N({\rm
  H_2})$ appears to be counter-rotating, possibly due to a satellite
(upper DLA data point).  Comparing the ISM gas-phase metallicity and
the metallicity of the DLA shows a possible gas metallicity gradient
of $\Delta\,Z/\Delta\,R=-0.02$ dex\,kpc$^{-1}$ out to $R\approx 14$
kpc.

The bottom row of Fig.~\ref{fig:outskirts} presents a DLA at $d=38$
kpc from an edge-on disk at $z=0.525$.  A strong velocity shear is
also seen along the disk of this $L_*$ galaxy.  Because the QSO
sightline occurs along the extended edge-on disk, Eq.~(3) and (4)
directly lead to $R\approx d$ and $v\approx v_{\rm obs}$ for this
system.  This DLA galaxy presents a second example for galaxies with
an extended rotating disk out to $\approx 40$\,kpc.  At the same time,
the deep $r$-band image (lower-left panel) from {\it HST} suggests
that the disk is warped near the QSO sightline, which is also
reflected by the presence of a disturbed rotation velocity at $R>5$
kpc (bottom-centre panel).  The metallicity measured in the gas phase
(bottom-right panel) displays a similar gradient\index{metallicity gradient} of $\Delta\,Z/\Delta\,R=-0.02$ dex\,kpc$^{-1}$ to the
galaxy at the top, which is also comparable to what is seen in the ISM
of nearby disk galaxies (e.g., \citealt{Zaritsky:1994, vanZee:1998,
  Sanchez:2014}).  A declining gas-phase metallicity from the inner
ISM to neutral gas at larger distances appears to hold for most DLA
galaxies at $z\lesssim  1$ and the declining trend continues into ionized
halo gas traced by strong LLS of $N(\HI)=10^{19-20}\,\cmjj$ (e.g.,
\citealt{Peroux:2016}).

At $z>2$, spatially resolved observations of ISM gas kinematics
become significantly more challenging, because the effective radii of
$L_*$ galaxies are typically $r_{\rm e}=1-3$\,kpc (e.g., \citealt{Law:2012}),
corresponding to $\lesssim  0.3''$, and smaller for fainter or lower-mass
objects.  Star-forming regions in these distant galaxies are barely
resolved in ground-based, seeing-limited observations (e.g.,
\citealt{Law:2007, Forster:2009, Wright:2009}).  Beam smearing can result
in significant bias in interpreting the observed velocity shear and
distributions of heavy elements (e.g., \citealt{Davies:2011,
  Wuyts:2016}).  However, accurate measurements can be obtained to
differentiate ISM metallicities of DLA galaxies from metallicities of
neutral gas beyond the star-forming regions.  Using the small sample
of known DLA galaxies at $z\gtrsim  2$, a metallicity gradient of
$\Delta\,Z/\Delta\,R=-0.02$ dex\,kpc$^{-1}$ is also found in these
distant star-forming galaxies (\citealt{Christensen:2014,
  Jorgenson:2014M}).

\section{The Star Formation Relation in the Early Universe}
\label{sec:sfr}

\begin{figure}
\begin{center}
\includegraphics[scale=0.45]{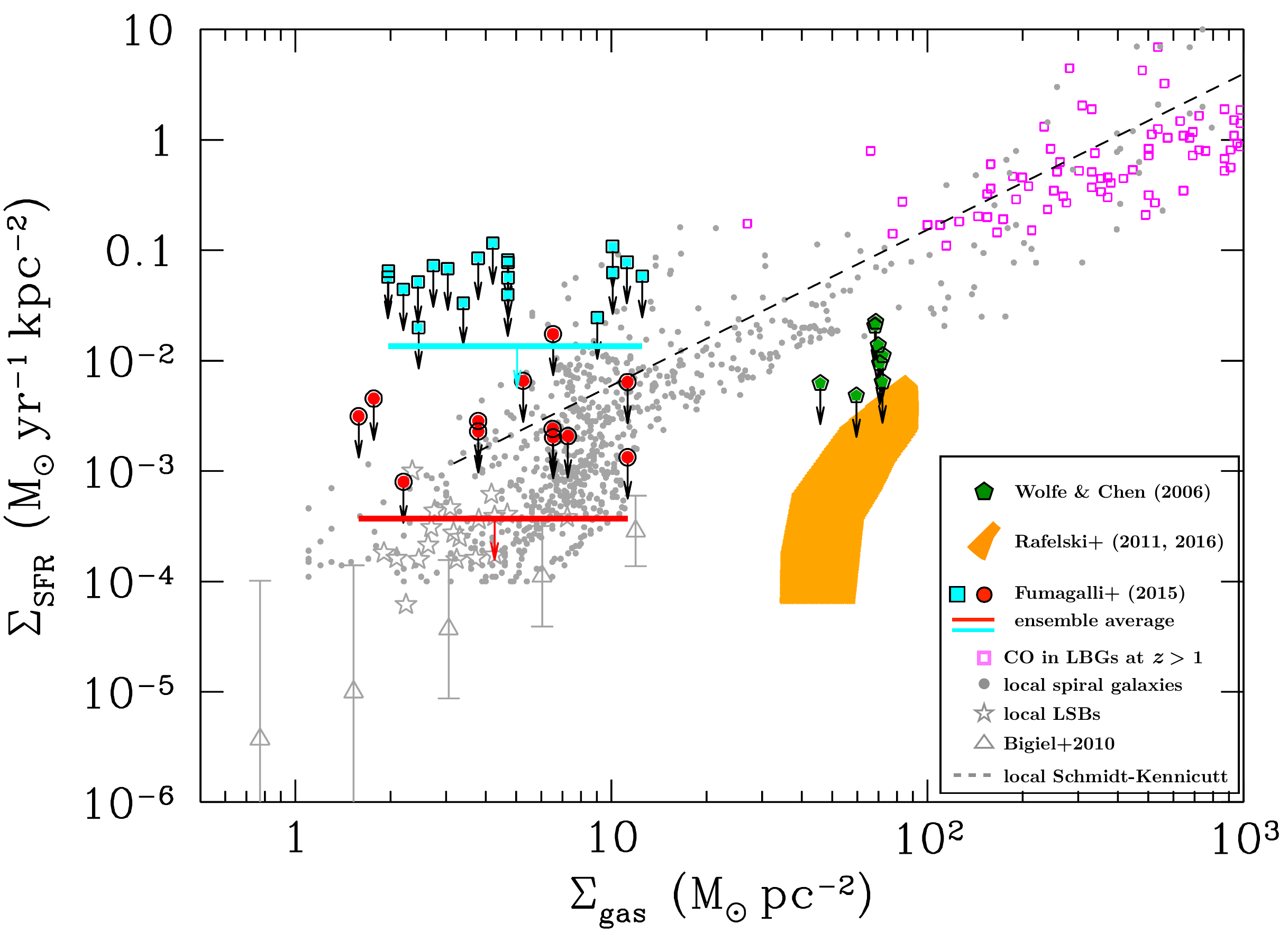}
\end{center}
\caption{The global star formation relation observed in nearby
  galaxies and at high redshifts.  The correlation between the SFR per unit area ($\Sigma_{\rm SFR}$) and the total
  surface gas mass density ($\Sigma_{\rm gas}$), combining both atomic
  (\HI) and molecular (H$_2$) for nearby spiral and starburst galaxies
  are shown in small filled circles (\citealt{Kennicutt:1998s,
    Gracia:2008, Leroy:2008}), together with the best-fit
  Schmidt-Kennicutt relation\index{Schmidt-Kennicutt relation} shown
  by the dashed line (\citealt{Kennicutt:1998s}).  A reduced star
  formation efficiency is observed both in low surface brightness
  galaxies and in the outskirts of normal spirals, which are shown in
  grey star symbols and open triangles, respectively
  (\citealt{Wyder:2009, Bigiel:2010}).  CO molecules have been detected
  in many massive starburst galaxies ($M_{\rm star}>2.5\times
  10^{10}\,\msol$) at $z=1-3$ (e.g., \citealt{Baker:2004, Genzel:2010,
    Tacconi:2013}), which occur at the high surface density regime of
  the global star formation relation (open squares).  In contrast,
  searching for {\it in situ} star formation in DLAs has revealed a
  reduced star formation efficiency in this metal-deficient gas.
  Specifically, green points and orange shaded area represent the
  constraints obtained from comparing the sky coverage of low surface
  brightness emission with the incidence of DLAs (\citealt{Wolfe:2006,
    Rafelski:2011, Rafelski:2016}).  Cyan squares and red circles
  represent the limits inferred from imaging searches of galaxies
  associated with individual DLAs, and the cyan and red bars represent
  the limiting $\Sigma_{\rm SFR}$ based on ensemble averages of the
  two samples (\citealt{Fumagalli:2015}).  The level of star formation
  observed in high-$N(\HI)$ DLAs (green pentagons and orange shaded
  area) is comparable to what is seen in nearby low surface brightness
  galaxies and in the outskirts of normal spirals.  See the main text
  for a detailed discussion}
\label{fig:sfrarea}       
\end{figure}

While direct identifications of galaxies giving rise to $z>2$ DLAs
have proven extremely challenging, critical insights into the star
formation relation in the early Universe can still be gained from
comparing the incidence of DLAs with the spatial distribution of star
formation rate (SFR) per unit area\index{star formation rate per unit area}
uncovered in deep imaging data (\citealt{Lanzetta:2002, Wolfe:2006}).
Specifically, the SFR per unit area ($\Sigma_{\rm
  SFR}$) is correlated with the surface mass density of neutral gas
($\Sigma_{\rm gas}$), following a Schmidt-Kennicutt
relation\index{Schmidt-Kennicutt relation} in nearby galaxies (e.g.,
\citealt{Schmidt:1959, Kennicutt:1998s}).  The global star formation
relation, $\Sigma_{\rm SFR}=2.5\times 10^{-4}\,(\Sigma_{\rm
  gas}/1\,\msol\,{\rm pc}^{-2})^{1.4}\,\msol\,{\rm yr}^{-1}\,{\rm
  kpc}^{-2}$ (dashed line in Fig.~\ref{fig:sfrarea}), is established
using a sample of local spiral galaxies and nuclear starbursts (solid
grey points in Fig.~\ref{fig:sfrarea}) over a broad range of
$\Sigma_{\rm gas}$, from $\Sigma_{\rm gas}\approx 10\,\msol\,{\rm
  pc}^{-2}$ to $\Sigma_{\rm gas}\approx 10^4\,\msol\,{\rm pc}^{-2}$.

Empirical constraints for a Schmidt-Kennicutt relation at high
redshifts require observations of the neutral gas content in
star-forming galaxies.  Although observations of individual galaxies
in \HI\ emission remain out of reach, the sample of $z=1-3$ galaxies
with resolved CO maps is rapidly growing (e.g., \citealt{Baker:2004,
  Genzel:2010, Tacconi:2013}).  The observed $\Sigma_{\rm SFR}$ versus
$\Sigma_{\rm molecular}$ for the high-redshift CO detected sample is
shown in open squares in Fig.~\ref{fig:sfrarea}, which occur at high
surface densities of $\Sigma_{\rm molecular}\gtrsim  100\,\msol\,{\rm
  pc}^{-2}$.  Considering only $\Sigma_{\rm molecular}$ is appropriate
for these galaxies, because locally it has been shown that at this
high surface density regime molecular gas dominates (e.g.,
\citealt{Martin:2001, Wong:2002, Bigiel:2008}).  In contrast, DLAs probe
neutral gas with $N(\HI)$ ranging from $N(\HI)=2\times 10^{20}\,\cmjj$
to $N(\HI)\approx 5\times 10^{22}\,\cmjj$.  The range in $N(\HI)$
corresponds to a range in surface mass density of atomic gas from
$\Sigma_{\rm atomic}\approx 2\,\msol\,{\rm pc}^{-2}$ to $\Sigma_{\rm
  atomic}\gtrsim  200\,\msol\,{\rm pc}^{-2}$, which is comparable to the
global average of total neutral gas surface mass density in local disk
galaxies (e.g., Fig.~\ref{fig:sfrarea}).  Therefore, DLAs offer an
important laboratory for investigating the star formation relation in
the distant Universe, and direct constraints can be obtained from
searches of {\it in situ} star formation in DLAs.

In principle, the Schmidt-Kennicutt relation can be rewritten in terms
of $N(\HI)$ for pure atomic gas following 
\begin{equation}
\Sigma_{\rm SFR}=K\times[N(\HI)/N_0]^\beta\ \ \ \msol\,{\rm yr}^{-1}\,{\rm kpc}^{-2},
\end{equation} 
which is justified for regions probed by DLAs with a low molecular gas
content (see Sect.~\ref{sec:DLAstats} and Fig.~\ref{fig:dla}d).
For reference, the local Schmidt-Kennicutt relation has $K=2.5\times
10^{-4}\,\msol\,{\rm yr}^{-1}\,{\rm kpc}^{-2}$, $\beta=1.4$, and
$N_0=1.25\times 10^{20}\,\cmjj$ for a pure atomic hydrogen gas.
Following Eq.~(5), the $N(\HI)$ distribution function,
$f_{N(\HI)}$ (e.g., Fig.~\ref{fig:dla}a), can then be expressed in
terms of the $\Sigma_{\rm SFR}$ distribution function, $h(\Sigma_{\rm
  SFR})$\index{star formation rate per unit area, distribution function}, which is the projected proper area per
$d\Sigma_{\rm SFR}$ interval per comoving volume
(\citealt{Lanzetta:2002}).  The $\Sigma_{\rm SFR}$ distribution function
$h(\Sigma_{\rm SFR})$ is related to $f_{N(\HI)}$ according to
$h(\Sigma_{\rm SFR})\,d\Sigma_{\rm SFR}=(H_0/c)\,f_{N(\HI)}\,dN(\HI)$.

This exercise immediately leads to two important observable
quantities.  First, the sky covering fraction ($C_{\rm A}$)\index{covering fraction} of star-forming regions in the redshift range,
$[z_1,z_2]$, with an observed SFR per unit area in the
interval of $\Sigma_{\rm SFR}$ and $\Sigma_{\rm SFR}+d\Sigma_{\rm
  SFR}$ is determined following
\begin{equation}
C_{\rm A}[\Sigma_{\rm SFR}|N(\HI)] = \displaystyle\int_{z_1}^{z_2} \frac{c\,(1+z)^2}{H(z)} h(\Sigma_{\rm SFR})\,d\Sigma_{\rm SFR}\,dz,
\end{equation}
where $c$ is the speed of light and $H(z)$ is the Hubble expansion
rate.  Equation~(6) is equivalent to $f_{N(\HI)}dN(\HI)dX$, where
$dX\equiv (1+z)^2\,H_0/H(z)\,dz$ is the comoving absorption
pathlength.  In addition, the first moment of $h(\Sigma_{\rm SFR})$
leads to the comoving SFR density
(\citealt{Lanzetta:2002, Hopkins:2005}),
\begin{equation}
\dot{\rho}_*(>\Sigma_{\rm SFR}^{\rm min})=\int_{\Sigma_{\rm SFR}^{\rm min}}^{\Sigma_{\rm SFR}^{\rm max}}\Sigma_{\rm SFR}h(\Sigma_{\rm SFR})\,d\Sigma_{\rm SFR}.
\end{equation}  
Constraints on the star formation relation at high redshift, namely
$K$ and $\beta$ in Eq.~(5), can then be obtained by comparing
$f_{N(\HI)}$-inferred $C_{\rm A}$ and $\dot{\rho}_*$ with results from
searches of low surface brightness emission in deep galaxy survey
data.  Furthermore, estimates of missing light in low surface
brightness regions can also be obtained using Eq.~(7) (e.g.,
\citealt{Lanzetta:2002, Rafelski:2011}).

In practice, Eq.~(5) is a correct representation only if disks
are not well formed and a spherical symmetry applies to the DLAs.  For
randomly oriented disks, corrections for projection effects are
necessary and detailed formalisms are presented in \cite{Wolfe:2006}
and \cite{Rafelski:2011}.  In addition, the inferred surface
brightness of {\it in situ} star formation in the DLA gas is extremely
low after accounting for the cosmological surface brightness dimming.
At $z=2-3$, only DLAs at the highest-$N(\HI)$ end of $f_{N(\HI)}$ are
expected to be visible in ultra-deep imaging data
(cf.\ \citealt{Lanzetta:2002, Wolfe:2006}).  For example, DLAs of
$N(\HI)>1.6\times 10^{21}\,\cmjj$ at $z\approx 3$ are expected to have
$V$-band (corresponding roughly to rest-frame 1500\,\AA\ at $z=3$)
surface brightness $\mu_V\lesssim  28.4$ mag arcsec$^{-2}$, assuming the
local Schmidt-Kennicutt relation.  The expected low surface brightness
of UV photons from young stars in high-redshift DLAs dictates the
galaxy survey depth necessary to uncover star formation associated
with the DLA gas.  At $N(\HI)>1.6\times 10^{21}\,\cmjj$, roughly 3\%
of the sky ($C_{\rm A}\approx 0.03$) is expected to be covered by extended
low surface brightness emission of $\mu_V\lesssim  28.4$ mag arcsec$^{-2}$.
For comparison, the sky covering fraction of luminous starburst
galaxies at $z=2-3$ is less than 0.1\%.

Available constraints for the star formation efficiency at $z=1-3$ are
shown in colour symbols in Fig.~\ref{fig:sfrarea}.  Specifically, the
Hubble Ultra Deep Field (HUDF; \citealt{Beckwith:2006})\index{Hubble Ultra Deep Field} $V$-band image offers sufficient depth for
detecting objects of $\mu_V\approx 28.4$ mag arcsec$^{-2}$.  Under the
assumption that DLAs originate in regions distinct from known
star-forming galaxies, an exhaustive search for extended low surface
brightness emission in the HUDF has uncovered only a small number of
these faint objects, far below the expectation from applying the local
Schmidt-Kennicutt relation for DLAs of $N(\HI)>1.6\times
10^{21}\,\cmjj$ following Eq.~(6).  Consequently, matching the
observed limit on $\dot{\rho_*}$ from these faint objects with
expectations from Eq.~(7) has led to the conclusion that the star
formation efficiency in metal-deficient atomic gas is more than
$10\times$ lower than expectations from the local Schmidt-Kennicutt
relation (\citealt{Wolfe:2006}; green pentagons in Fig.~\ref{fig:sfrarea}).

On the other hand, independent observations of DLA galaxies at $z=2-3$
have suggested that these absorbers are associated with typical
star-forming galaxies at high redshifts.  These include a comparable
clustering amplitude of DLAs and these galaxies (e.g.,
\citealt{Cooke:2006}), the findings of a few DLA galaxies with mass and
SFR comparable to luminous star-forming galaxies found
in deep surveys (e.g., \citealt{Moller:2002, Moller:2004,
  Christensen:2007}), and detections of a DLA feature in the ISM of
star-forming galaxies (e.g., \citealt{Pettini:2002, Chen:2009GRB,
  Dessauges:2010}).  If DLAs originate in neutral gas around known
star-forming galaxies, then these luminous star-forming galaxies
should be more spatially extended than has been realized.  Searches
for low surface brightness emission in the outskirts of these galaxies
based on stacked images have indeed uncovered extended low surface
brightness emission out to more than twice the optical extent of a
single image.  However, repeating the exercise of computing the
cumulative $\dot{\rho}_*$ from Eq.~(7) has led to a similar
conclusion that the star formation efficiency\index{star formation efficiency} is more than $10\times$ lower in metal-deficient atomic
gas at $z=1-3$ than expectations from the local Schmidt-Kennicutt
relation (\citealt{Rafelski:2011, Rafelski:2016}).  The results are shown
as the orange shaded area in Fig.~\ref{fig:sfrarea}).  In addition,
the amount of missing light in the outskirts of these luminous
star-forming galaxies is found to be $\approx 10$\% of what is
observed in the core (\citealt{Rafelski:2011}).

At the same time, imaging searches of individual DLA galaxies have
been conducted for $\approx 30$ DLAs identified along QSO sightlines
that have high-redshift LLS serving as a natural coronograph to block
the background QSO glare, improving the imaging depth in areas
immediate to the QSO sightline (\citealt{Fumagalli:2015}).  These
searches have yielded only null results, leading to upper limits on
the underlying surface brightness of the DLA galaxies (cyan squares
and red circles in Fig.~\ref{fig:sfrarea}).  While the survey depth
is not sufficient for detecting associated star-forming regions in
most DLAs in the survey sample of \cite{Fumagalli:2015} based on the
local Schmidt-Kennicutt relation, the ensemble average is beginning to
place interesting limits (cyan and red arrows).

The lack of {\it in situ} star formation in DLAs may not be surprising
given the low molecular gas content.  In the local Universe, it is
understood that the Schmidt-Kennicutt relation is driven primarily by
molecular gas mass ($\Sigma_{\rm molecular}$), while the surface
density of atomic gas ($\Sigma_{\rm atomic}$) ``saturates'' at $\sim
10\,\msol\,{\rm pc}^{-2}$ beyond which the gas transitions into the
molecular phase (e.g., \citealt{Martin:2001, Wong:2002, Bigiel:2008}).
As described in Sect.~\ref{sec:DLAstats} and Fig.~\ref{fig:dla}d,
the transitional surface density from atomic to molecular is
metallicity dependent.  Therefore, the low star formation efficiency
observed in DLA gas can be understood as a metallicity-dependent
Schmidt-Kennicutt relation.  This is qualitatively consistent with the
observed low $\Sigma_{\rm SFR}$ in nearby low surface brightness
galaxies (e.g., \citealt{Wyder:2009}; star symbols in Fig.~\ref{fig:sfrarea}) and in the outskirts of normal spirals (e.g.,
\citealt{Bigiel:2010}; open triangles in Fig.~\ref{fig:sfrarea}), where
the ISM is found to be metal-poor (e.g., \citealt{McGaugh:1994,
  Zaritsky:1994, Bresolin:2012}).  Numerical simulations incorporating
a metallicity dependence in the H$_2$ production rate have also
confirmed that the observed low star formation efficiency in DLAs can
be reproduced in metal-poor gas (e.g., \citealt{Gnedin:2010}).  

A metallicity-dependent Schmidt-Kennicutt relation has wide-ranging
implications in extragalactic research, from the physical origin of
DLAs at high redshifts, to star formation and chemical enrichment
histories in different environments, and to detailed properties of
distant galaxies such as morphologies, sizes, and cold gas content.
It is clear from Fig.~\ref{fig:sfrarea} that there exists a
significant gap in the gas surface densities, between $\Sigma_{\rm
  gas}\approx 10\, \msol\,{\rm pc}^{-2}$ probed by these direct DLA
galaxy searches and $\Sigma_{\rm gas}\approx 100\, \msol\,{\rm
  pc}^{-2}$ probed by CO observations of high-redshift starburst
systems (open squares in Fig.~\ref{fig:sfrarea}).  Continuing
efforts targeting high-$N(\HI)$ DLAs (and therefore high $\Sigma_{\rm
  gas}$) at sufficient imaging depths are expected to place critical
constraints on the star formation relation in low-metallicity
environments at high redshifts.  Similarly, spatially resolved maps of
star formation and neutral gas at $z>1$ to mean surface densities of
$\Sigma_{\rm SFR}< 0.1\,\msol\,{\rm yr}^{-1}\,{\rm kpc}^{-2}$ and
$\Sigma_{\rm atomic, molecular}\approx 10-100\,\msol\,{\rm pc}^{-2}$
will bridge the gap of existing observations and offer invaluable
insights into the star formation relation in different environments.

\section{From Neutral ISM to the Ionized Circumgalactic Medium}
\label{sec:cgm}

Beyond the neutral ISM, strong \lya\ absorbers of $N(\HI)\approx
10^{14-20}\,\cmjj$ and associated metal-line absorbers offer a
sensitive probe of the diffuse circumgalactic medium (CGM)\index{circumgalactic medium} to projected distances $d\approx
100-500$\,kpc (e.g., Fig.~\ref{fig:halomap}). But because the
circumgalactic gas is significantly more ionized in the LLS and
lower-$N(\HI)$ regime, measurements of its ionization state and
metallicity bear considerable uncertainties and should be interpreted
with caution.

Several studies have attempted to constrain the ionization state and
metallicity of the CGM by considering the relative abundances of
different ions at low- and high-ionization states (e.g.,
\citealt{Savage:2002, Stocke:2006}).  For example, attributing observed
\OVI\ absorbers\index{O\,{\sc vi} absorbers} to cool ($T\sim 10^4$\,K),
photo-ionized gas irradiated by the metagalactic ionizing radiation
field, the observed column density ratios between \OVI\ and
low-ionization transitions (such as \CIII\ and \CIV) require extremely
low gas densities of $n_{\rm H}\sim 10^{-5}\,{\rm cm}^{-3}$.
Combining the inferred low gas density with observed $N(\OVI)$, which
are typically $\gtrsim  10^{14.5}\,\cmjj$ in galactic haloes (e.g.,
\citealt{Tumlinson:2011}), leads to a moderate gas metallicity of $\gtrsim 
1/10$ Solar and unphysically large cloud sizes of $l_c\sim 1$ Mpc
(e.g., \citealt{Tripp:2001, Savage:2002, Stocke:2006})\footnote{For
  comparison, the sizes of extended HVC complexes at $d\sim 10$\,kpc
  from the MW disk are a few to 15\,kpc across (e.g.,
  \citealt{Putman:2012}).  HVCs at larger distances are
  found to be more compact, $\lesssim  2$\,kpc (e.g., \citealt{Westmeier:2008,
    Lockman:2012, Giovanelli:2013}). }.  Excluding \OVI\ due to
possible origins in shocks or turbulent mixing layers (e.g.,
\citealt{Heckman:2002OVI}) and considering only relative abundances of
low-ionization species increases estimated gas densities to $n_{\rm
  H}\sim 10^{-4}-10^{-3}\,{\rm cm}^{-3}$.  The inferred cloud
sizes\index{cloud size} remain large with $l_c\sim 10-100$\,kpc, in
tension with what is observed locally for the HVCs.  The implied
thermal pressures in the cool gas phase are still two orders of
magnitude lower than what is expected from pressure equilibrium with a
hot ($T\approx 10^6$\,K) medium (e.g., \citealt{Stocke:2013, Werk:2014}),
indicating that these clouds would be crushed quickly.  Considering
non-equilibrium conditions (e.g., \citealt{Gnat:2007, Oppenheimer:2013})
and the presence of local ionizing sources may help alleviate these
problems (e.g., \citealt{Cantalupo:2010}), but the systematic
uncertainties are difficult to quantify.

Nevertheless, exquisite details concerning extended halo gas have been
learned over the past decade based on various samples of close galaxy
and background QSO pairs.  Because luminous QSOs are rare, roughly one
QSO of $g\lesssim  18$ mag per square degree (e.g., \citealt{Richards:2006}),
absorption-line studies of the CGM against background QSO light have
been largely limited to one probe per galaxy.  Only in a few cases are
multiple QSOs found at $d\lesssim  300$\,kpc from a foreground galaxy (e.g.,
\citealt{Norman:1996, Keeney:2013, Davis:2015, Lehner:2015, Bowen:2016})
for measuring coherence in spatial distribution and kinematics of
extended gas around the galaxy.  All of these cases are in the local
Universe, because the relatively large angular extent of these
galaxies on the sky increases the probability of finding more than one
background QSO.  This local sample has now been complemented with new
studies, utilizing multiply lensed QSOs and close projected QSO pairs,
which provide spatially resolved CGM absorption properties for a
growing sample of galaxies at intermediate redshifts (e.g.,
\citealt{Chen:2014, Rubin:2015, Zahedy:2016}).

\begin{figure}
\begin{center}
\includegraphics[scale=0.4]{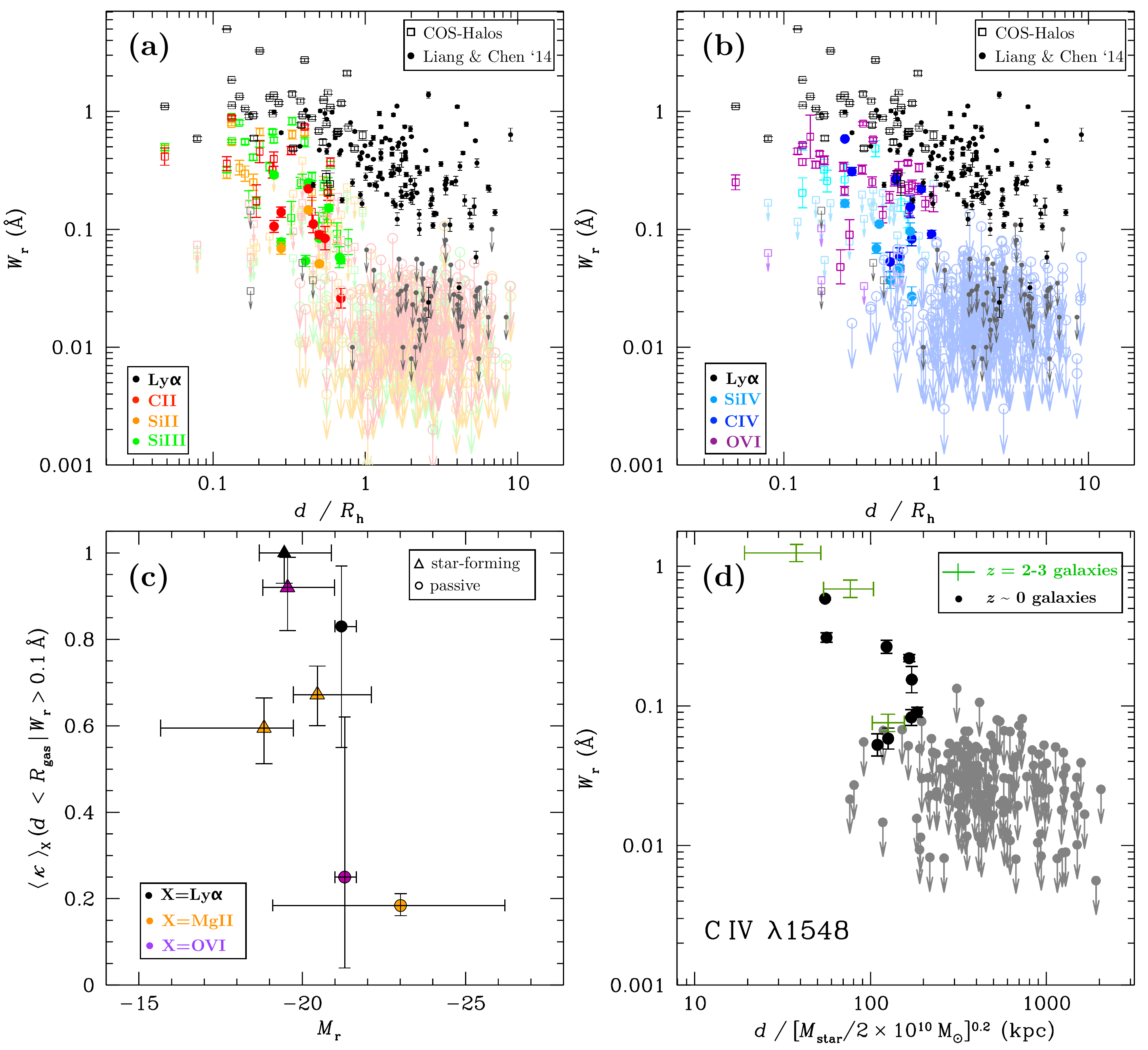}
\end{center}
\caption{Observed absorption properties of halo gas around galaxies.
  The {\it top} panels display the radial profiles\index{circumgalactic medium, radial profiles} of rest-frame absorption equivalent width ($W_{\rm r}$) versus
  halo-radius $R_{\rm h}$-normalized projected distance for different
  absorption transitions.  Low-ionization transitions are presented in
  panel ({\it a}) and high-ionization transitions in panel ({\it b}).  \lya\ data
  points are presented in both panels for cross-comparison.  The
  galaxy sample includes 44 galaxies at $z\approx 0.25$ from the
  COS-Halos project (open squares; \citealt{Tumlinson:2011,
    Tumlinson:2013, Werk:2013}) and $\sim 200$ galaxies at $z\approx
  0.04$ from public archives (circles; \citealt{Liang:2014}), for which
  high-quality, ultraviolet QSO spectra are available for constraining
  the presence or absence of multiple ions in individual haloes.
  Different transitions are colour-coded to highlight the differences
  in their spatial distributions.  For transitions that are not
  detected, a 2-$\sigma$ upper limit is shown by a downward arrow.  No
  heavy ions are found beyond $d=R_{\rm h}$, while \lya\ continues to be
  seen to larger distances.  Panel ({\it c}) displays the ensemble average
  of gas covering fraction ($\langle\kappa\rangle$)\index{covering fraction} as a function of absolute $r$-band magnitude ($M_{\rm r}$) for
  \lya\ (black symbols), \MgII\ (orange), and \OVI\ (purple).
  Star-forming galaxies (triangles) on average are fainter and exhibit
  higher covering fractions of hydrogen and chemically enriched gas
  probed by both low- and high-ionization species than passive
  galaxies (circles).  Measurements of \lya- and \OVI-absorbing gas
  are based on COS-Halos galaxies for $R_{\rm gas}=R_{\rm h}$.  Measurements
  of \MgII-absorbing gas are based on $\approx 260$ star-forming
  galaxies at $z\approx 0.25$ (\citealt{Chen:2010}), and $\sim 38000$
  passive luminious red galaxies at $z\approx 0.5$ (\citealt{Huang:2016})
  for $R_{\rm gas}=R_{\rm h}/3$.  Panel ({\it d}) illustrates the apparent
  constant nature of mass-normalized radial profiles of CGM absorption
  since $z\approx 3$ (e.g., \citealt{Chen:2012, Liang:2014}).  The
  high-redshift observations are based on mean \CIV\ absorption in
  stacked spectra of $\sim 500$ starburst galaxies with a mean stellar
  mass and dispersion of $\langle\,\log\,M_{\rm star}\,\rangle=9.9\pm
  0.5$ (\citealt{Steidel:2010}), and the low-redshift observations are
  for $\sim 200$ individual galaxies with $\langle\,\log\,M_{\rm
    star}\,\rangle=9.7\pm 1.1$ and modest SFR (\citealt{Liang:2014}) }
\label{fig:cgm}       
\end{figure}

With one QSO probe per halo, a two-dimensional map of CGM absorption
properties can be established based on an ensemble average of a large
sample of QSO-galaxy pairs ($N_{\rm pair}\sim 100-1000$).  Fig.\ref{fig:cgm} summarizes some of the observable quantities of the CGM.
First, panels (a) and (b) at the top display the radial
profiles\index{circumgalactic medium, radial profiles} of rest-frame absorption
equivalent width ($W_{\rm r}$) for different absorption transitions,
including hydrogen \lya, low-ionization \CII\ and \SiII,
intermediate-ionization \SiIII, \SiIV, and \CIV, and high-ionization
\OVI\ absorption transitions, colour-coded in black, red, orange,
green, blue, magenta, and dark purple, respectively.  For transitions
that are not detected, a 2-$\sigma$ upper limit is shown as a downward
arrow.  Because of the large number of data points, the upper limits
are shown in pale colours for clarity.  The galaxy sample includes 44
galaxies at $z\approx 0.25$ from the COS-Halos project (open squares;
\citealt{Tumlinson:2011, Tumlinson:2013, Werk:2013}) and $\sim 200$
galaxies at $z\approx 0.04$ from public archives (circles;
\citealt{Liang:2014}), for which high-quality, ultraviolet QSO spectra
are available for constraining the presence or absence of multiple
ions in individual haloes.  These galaxies span four decades in
total stellar mass, from $M_{\rm star}\approx 10^7\,\msol$ to $M_{\rm
  star}\approx 10^{11}\,\msol$, and a wide range in 
SFR, from ${\rm SFR}<0.1\,\msol\,{\rm yr}^{-1}$ to ${\rm
  SFR}>10\,\msol\,{\rm yr}^{-1}$.  Diffuse gas is observed beyond
$d=50$\,kpc around distant galaxies, extending the detection limit of
\HI\ gas in inner galactic haloes from 21\,cm observations (e.g., Fig.~\ref{fig:halomap}) to lower column density gas at larger distances and
higher redshifts.

While $W_{\rm r}$ is typically found to decline steadily with increasing $d$
for all transitions (e.g., \citealt{Chen:2012,Werk:2014}), the scatters
are large.  Including the possibility that more massive haloes have
more spatially extended halo gas, the halo radius $R_{\rm h}$-normalized
$W_{\rm r}$-$d$ distribution indeed displays substantially reduced scatters
in the radial profiles shown in panels (a) and (b) of Fig.~\ref{fig:cgm}.  A reduced scatter in the $R_{\rm h}$-normalized $W_{\rm r}$-$d$
distribution indicates that galaxy mass plays a dominant role in
driving the extent of halo gas.  In addition, it also confirms that
accurate associations between absorbers and absorbing galaxies have
been found for the majority of the systems.

A particularly interesting feature in Fig.~\ref{fig:cgm} is a
complete absence of heavy ions beyond $d=R_{\rm h}$, while detections of
\lya\ continue to larger distances.  The absence of heavy ions at
$d>R_{\rm h}$, which is observed for a wide range of ionization states,
strongly indicates that chemical enrichment is confined within
individual galaxy haloes.  This finding applies to both low-mass dwarfs
and massive galaxies.  However, it should also be noted that heavy
ions are observed beyond $R_{\rm h}$ for galaxies with close neighbours
(e.g., \citealt{Borthakur:2013, Johnson:2015}), suggesting that
environmental effects play a role in distributing heavy elements
beyond the enriched gaseous haloes of individual galaxies.  Comparing
panels (a) and (b) of Fig.~\ref{fig:cgm} also shows that within
individual galaxy haloes, a global ionization gradient\index{ionization gradient} is seen with more highly ionized gas detected at larger
distances.  For instance, the observed $W_{\rm r}$ declines to $<0.1$
\AA\ at $d\approx 0.5\,R_{\rm h}$ for \CII\ and \SiII, while \CIV\ and
\OVI\ absorbers of $W_{\rm r}>0.1$\,\AA\ continue to be found beyond
$0.5\,R_{\rm h}$.

The observed $W_{\rm r}$ versus $d$ (or $d/R_{\rm h}$) based on a blind survey of
absorption features in the vicinities of known galaxies also enables
measurements of gas covering fraction\footnote{A blind survey of
  absorption features around known galaxies differs fundamentally from
  a blind survey of galaxies around known absorbers (e.g.,
  \citealt{Kacprzak:2008}).  By design, a blind galaxy survey around
  known absorbers excludes transparent sightlines and does not provide
  the sample necessary for measuring the incidence and covering
  fraction of absorbing species.  In addition, because of limited
  survey depths, a blind galaxy survey is more likely to find more
  luminous members at larger $d$ that are correlated with the true
  absorbing galaxies which are fainter and closer to the QSO
  sightline, resulting in a significantly larger scatter in the $W_{\rm r}$
  versus $d$ distribution (e.g., \citealt{Kacprzak:2008,
    Nielsen:2013}).}.  The mean gas covering fraction
($\langle\kappa\rangle$) can be measured by a simple accounting of the
fraction of galaxies in an annular area displaying associated
absorbers with $W_{\rm r}$ exceeding some detection threshold $W_0$, and
uncertainties can be estimated based on a binomial distribution
function.  Dividing the sample into different projected distance bins,
it is clear from Fig.~\ref{fig:cgm}a and b that the fraction of
non-detections increases with increasing projected distance, resulting
in a declining $\langle\kappa\rangle$ with increasing $d$ for all
transitions observed (see also \citealt{Chen:2010, Werk:2014,
  Huang:2016}).

It is also interesting to examine how $\langle\kappa\rangle$ depends
on galaxy properties.  Figure~\ref{fig:cgm}c displays
$\langle\kappa\rangle$ observed within a fiducial gaseous radius
$R_{\rm gas}$ for star-forming (triangles) and passive (circles)
galaxies.  The measurements are made for \lya\ (black symbols),
\MgII\ (orange), and \OVI\ (purple) with a threshold of $W_0=0.1$\,\AA,
and shown in relation to the absolute $r$-band magnitude ($M_{\rm r}$).
Error bars represent the 68\% confidence interval.  The absolute
$r$-band magnitude is a direct observable of a galaxy and serves as a
proxy for its underlying total stellar mass.  Measurements of \lya-
and \OVI-absorbing gas are based on COS-Halos galaxies for $R_{\rm
  gas}=R_{\rm h}$ (see also \citealt{Johnson:2015} for a sample compiled from
the literature).  Measurements of \MgII-absorbing gas are based on
$\approx 260$ star-forming galaxies at $z\approx 0.25$
(\citealt{Chen:2010}, and $\sim 38000$ passive luminous red galaxies at
$z\approx 0.5$ (\citealt{Huang:2016}) for $R_{\rm gas}=R_{\rm h}/3$ (e.g.,
\citealt{Chen:2008}).  The larger sample sizes led to better constrained
$\langle\kappa\rangle$ for \MgII\ absorbing gas in galactic haloes.  In
general, star-forming galaxies on average are fainter, less massive,
and exhibit a higher covering fraction of chemically enriched gas than
passive galaxies (see also \citealt{Johnson:2015}).  At the same time,
the covering fraction of chemically enriched gas is definitely
non-zero around massive quiescent galaxies.

Comparing the radial profiles of CGM absorption at different redshifts
offers additional insights into the evolution history of the CGM,
which in turn helps distinguish between different models for chemical
enrichment in galaxy haloes.  The radial profiles of the CGM have been
found to evolve little since $z\sim 3$ (e.g., \citealt{Chen:2012}), even
though the star-forming properties in galaxies have evolved
significantly.  Figure~\ref{fig:cgm}d illustrates the apparent
constant nature of mass-normalized radial profiles of \CIV\ absorption
in galactic haloes (\citealt{Liang:2014}).  The high-redshift observations
are based on stacked spectra of $\sim 500$ starburst galaxies with a
mean stellar mass and dispersion of $\langle\,\log\,M_{\rm
  star}\,\rangle=9.9\pm 0.5$ (\citealt{Steidel:2010}) and a mean SFR of
$\langle\,{\rm SFR}\,\rangle\approx 30-60\,\msol\,{\rm yr}^{-1}$
(e.g., \citealt{Erb:2006SFR, Reddy:2012}).  The low-redshift galaxy
sample contains individual measurements of $\sim 200$ galaxies with
$\langle\,\log\,M_{\rm star}\,\rangle=9.7\pm 1.1$ and more quiescent
star-forming activities of $\langle\,{\rm SFR}\,\rangle\sim
1\,\msol\,{\rm yr}^{-1}$ (\citealt{Chen:2012, Liang:2014}).  The constant
mass-normalized CGM radial profiles between galaxies of very different
SFR indicate that mass (rather than SFR) is a dominant factor that
determines the CGM properties over a cosmic time interval.  This is
consistent with previous findings that CGM absorption properties
depend strongly on galaxy mass but only weakly on SFR (e.g.,
\citealt{Chen:2010M}), but at odds with popular models that attribute
metal-line absorbers to starburst-driven outflows (e.g.,
\citealt{Steidel:2010, Menard:2011}).

\begin{figure}
\begin{center}
\includegraphics[scale=0.48]{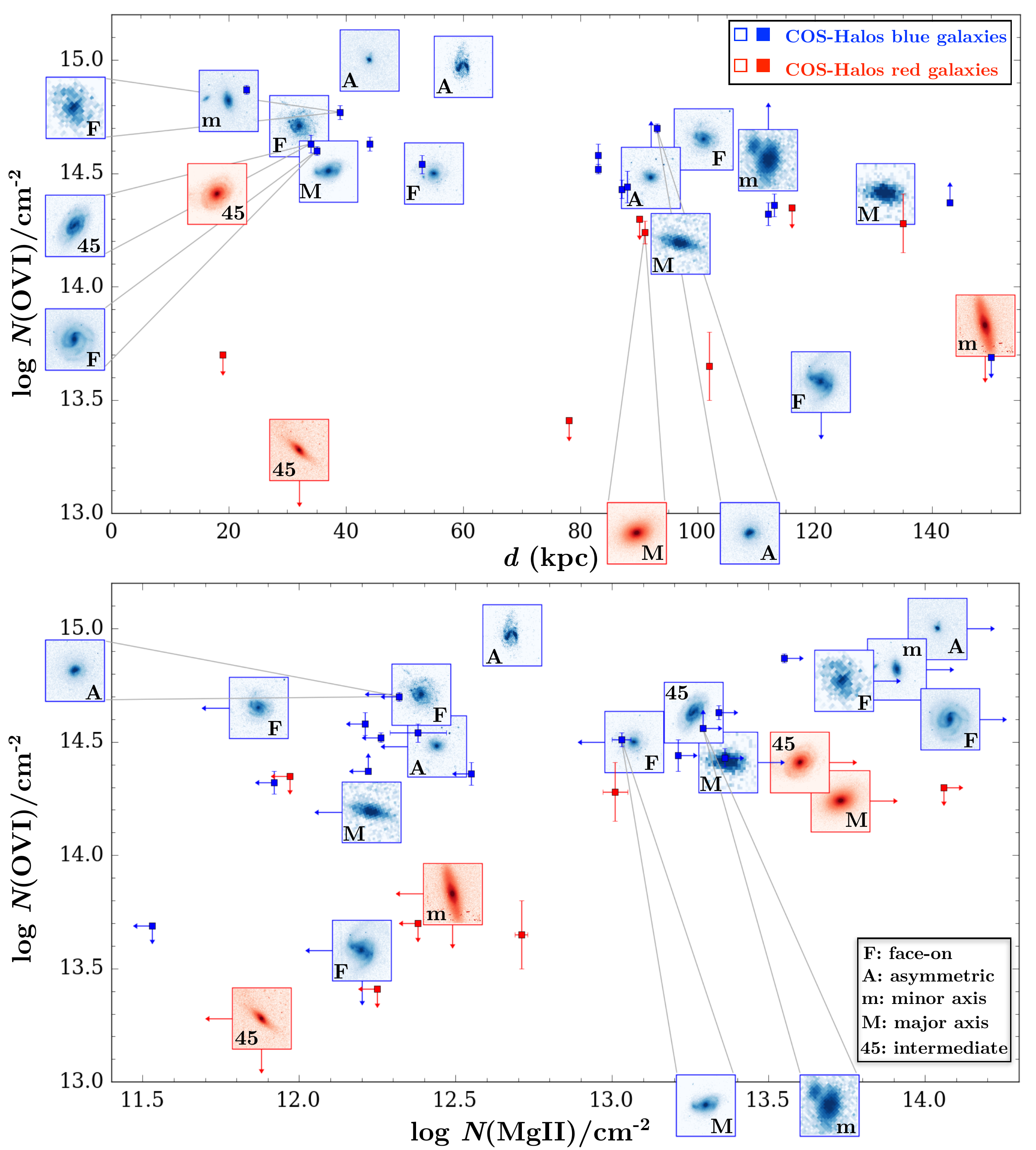}
\end{center}
\caption{Visual comparisons of the geometric alignment of galaxy major
  axis relative to the QSO sightline and the observed CGM absorption
  strength (by Rebecca Pierce).  {\it Top}: Observed \OVI\ column
  density\index{O\,{\sc vi} absorbers}, $N(\OVI)$, versus $d$ for COS-Halos
  star-forming (in blue) and passive (in red) galaxies
  (\citealt{Tumlinson:2011}).  {\it Bottom}: Comparisons of $N(\OVI)$ and
  $N(\MgII)$\index{Mg\,{\sc ii} absorbers} for the COS-Halos galaxies from
  \cite{Werk:2013}.  When spatially resolved images are available, the
  data points are replaced with an image panel of the absorbing
  galaxy.  Each panel is 25 proper kpc on a side, and is oriented such
  that the QSO sightline occurs on the y-axis at the corresponding
  \OVI\ column density of the galaxy.  Disk alignments cannot be
  determined for face-on galaxies (minor-to-major axis ratio $>0.7$)
  and galaxies displaying irregular/asymmetric morphologies, which are
  labeled ``F'' and ``A'', respectively.  Galaxies with the QSO
  located within $30^\circ$ of the minor axis are labeled 'm' in the
  lower-left corner, while galaxies with the QSO located within
  $30^\circ$ of the major axis are labeled 'M'.  Galaxies with the QSO
  sightline occuring intermediate ($30^\circ-60^\circ$) between the
  minor and major axis are labeled ``45''.  Downward arrows indicate
  2-$\sigma$ upper limits for non-detections, while upward arrows
  indicate saturated absorption lines.  The COS-Halos galaxy sample
  provides a unique opportunity to examine low- and high-ionization
  halo gas for the same galaxies at once.  Galaxies surrounded by
  \OVI\ and \MgII\ absorbing gas clearly exhibit a broad range both in
  morphology\index{galaxy morphology} and in disk orientation.  In
  addition, the observed $N(\MgII)$ displays a significantly larger
  scatter than $N(\OVI)$}
\label{fig:align}       
\end{figure}

A discriminating characteristic of starburst-driven
outflows\index{starburst outflows} is their distinctly non-spherical
distribution in galactic haloes in the presence of a well-formed
star-forming disk.  Specifically, galactic-scale outflows are expected
to travel preferentially along the polar axis where the gas
experiences the least resistance (e.g., \citealt{Heckman:1990}).  In
contrast, accretion of the IGM is expected to proceed along the disk
plane with $\lesssim  10$\% covering fraction on the sky (e.g.,
\citealt{Faucher:2011, Fumagalli:2011}).  Such azimuthal
dependence\index{azimuthal dependence} of the spatial distribution of
infalling and outflowing gas is fully realized in state-of-the-art
cosmological zoom-in simulations (e.g., \citealt{Shen:2013,
  Agertz:2015}).  Observations of $z\approx 0.7$ galaxies have shown
that at $d<50$\,kpc the mean \MgII\ absorption equivalent width within
$45^\circ$ of the minor axis is twice of the mean value found within
$45^\circ$ of the major axis, although such azimuthal dependence is
not observed at $d>50$\,kpc (\citealt{Bordoloi:2011}).  The observed
azimuthal dependence of the mean \MgII\ absorption strength is
qualitatively consistent with the expectation that these heavy ions
originate in starburst-driven outflows, and the lack of such azimuthal
dependence implies that starburst outflows are confined to the inner
halo of $d\lesssim  50$\,kpc.

Many subsequent studies have generalized this observed azimuthal
dependence at $d<50$\,kpc to larger distances and attributed absorbers
detected near the minor axis to starburst-driven outflows and those
found near the major axis to accretion (e.g., \citealt{Bouche:2012,
  Kacprzak:2015}).  However, a causal connection between the observed
absorbing gas and either outflows or accretion remains to be
established.  While gas metallicity may serve as a discriminator with
the expectation of starburst outflows being more metal-enriched
relative to the low-density IGM, uncertainties arise due to poorly
understood chemical mixing and metal transport (e.g.,
\citealt{Tumlinson:2006}).  Incidentally, a relatively strong
\MgII\ absorber has been found at $d\approx 60$\,kpc along the minor
axis of a starburst galaxy but the metallicity of the absorbing gas is
10 times lower than what is observed in the ISM
(\citealt{Kacprzak:2014}), highlighting the caveat of applying gas
metallicity as the sole parameter for distinguishing between accretion
and outflows.

Figure~\ref{fig:align} presents visual comparisons of the geometric
alignment of galaxy major axis relative to the QSO sightline and the
observed CGM absorption strength.  The figure at the top displays the
observed \OVI\ column density, $N(\OVI)$, versus $d$ for COS-Halos
galaxies at $z\approx 0.2$ (\citealt{Tumlinson:2013}).  The bottom figure
displays comparisons of $N(\OVI)$ and $N(\MgII)$ for these galaxies.
The absorption-line measurements are adopted from \cite{Werk:2013}.
When spatially resolved images are available, the data points are
replaced with an image panel of the absorbing galaxy.  Each panel is
25 proper kpc on a side, and is oriented such that the QSO sightline
falls on the y-axis at the corresponding $N(\OVI)$ of the galaxy.
The relative
alignment between galaxy major axis and the background QSO sightline
cannot be determined, if the galaxies are face-on with a
minor-to-major axis ratio $>0.7$ or if the galaxies display
irregular/asymmetric morphologies.  These galaxies are labeled ``F''
and ``A'', respectively.  For galaxies that clearly display a smooth
and elongated morphology, the orietation of the major axis can be
accurately measured.  Galaxies with the QSO located within $30^\circ$
of the minor axis are labeled 'm', while galaxies with the QSO located
within $30^\circ$ of the major axis are labeled 'M'.  Galaxies with
the QSO sightline occuring intermediate ($30^\circ-60^\circ$) between
the minor and major axis are labeled ``45''.  Star-forming galaxies
are colour-coded in blue, and passive galaxies in red.  Downward arrows
indicate 2-$\sigma$ upper limits for non-detections, while upward
arrows indicate saturated absorption lines.

While the COS-Halos sample is small, particularly when restricting to
those galaxies displaying a smooth, elongated morphology, it provides a unique
opportunity to examine low- and high-ionization halo gas for the same
galaxies at once.  Two interesting features are immediately clear in
Fig.~\ref{fig:align}.  First, galaxies surrounded by \OVI\ and
\MgII\ absorbing gas exhibit a broad range both in
morphology\index{galaxy morphology} and in star formation history,
from compact quiescent galaxies, to regular star-forming disks, and to
interacting pairs.  The diverse morphologies in \OVI\ and
\MgII\ absorbing galaxies illuminate the challenge and uncertainties
in characterizing their relative geometric orientation to the QSO
sightline based on azimuthal angle alone.  When considering only
galaxies with smooth and elongated (minor-to-major axis ratio $<0.7$)
morphologies, no clear dependence of $N(\OVI)$ or $N(\MgII)$ on galaxy
orientation is found.  Specifically, nine star-forming galaxies
displaying strong \OVI\ absorption at $d<80$\,kpc ($\log\,N(\OVI)$)
have spatially resolved images available.  Two of these galaxies
display disturbed morphologies and four are nearly face-on.  The
remaining three galaxies have the inclined disks oriented at
$0^\circ$, $45^\circ$, and $90^\circ$ each.  For passive red galaxies,
two have spatially resolved images available and both are elongated
and aligned at $\approx 45^\circ$ from the QSO sightline.  One
displays an associated strong \OVI\ absorber and the other has no
corresponding \OVI\ detections.  At $d>80$\,kpc, the morphology
distribution is similar to those at smaller distances.  No strong
dependence is found between the presence or absence of a strong
\OVI\ absorber and the galaxy orientation.

In addition, while the observed $N(\OVI)$ at $d<100\,kpc$ appears to be
more uniformly distributed with a mean and scatter of
$\log\,N(\OVI)=14.5\pm 0.3$ (\citealt{Tumlinson:2011}), the observed
$N(\MgII)$ displays a significantly larger scatter.  Specifically, the
face-on galaxy at $d\approx 32$\,kpc with an associated \OVI\ absorber
of $\log\,N(\OVI)\approx 14.7$ does not have an associated
\MgII\ absorber detected to a limit of $\log\,N(\MgII)\approx 12.4$.
Two quiescent galaxies at $z\approx 20$ and 90\,kpc (red panels)
exhibit saturated \MgII\ absorption of $\log\,N(\MgII)>13.5$ and
similarly strong \OVI\ of $\log\,N(\OVI)\approx 14.3$.  A small
scatter implies a more uniformly distributed medium, while a large
scatter implies a more clumpy nature of the absorbing gas or a larger
variation between different galaxy haloes.  Such distinct spatial
distributions between low- and high-ionization gas further highlight
the complex nature of the chemically enriched CGM, which depends on
more than the geometric alignment of the galaxies.  A
three-dimensional model of gas kinematics that takes full advantage of
the detailed morphologies and star formation history of the galaxies
is expected to offer a deeper understanding of the physical origin of
chemically enriched gas in galaxy haloes (e.g., \citealt{Gauthier:2012,
  Chen:2014, Diamond:2016}).

\section{Summary}
\label{sec:summary}

QSO absorption spectroscopy provides a sensitive probe of both neutral
medium and diffuse ionized gas in the distant Universe.  It extends
21\,cm maps of gaseous structures around low-redshift galaxies both to
lower gas column densities and to higher redshifts.  Specifically,
DLAs of $N(\HI)\gtrsim  2\times 10^{20}\,\cmjj$ probe neutral gas in the
ISM of distant star-forming galaxies, LLS of $N(\HI)>10^{17}\,\cmjj$
probe optically thick HVCs and gaseous streams in and around galaxies,
and strong \lya\ absorbers of $N(\HI)\approx 10^{14-17}\,\cmjj$ and
associated metal-line absorption transitions, such as \MgII, \CIV, and
\OVI, trace chemically enriched, ionized gas and starburst outflows.
Over the last decade, an unprecedentedly large number of $\sim 10000$
DLAs have been identified along random QSO sightlines to provide
robust statistical characterizations of the incidence and mass density
of neutral atomic gas at $z\lesssim  5$.  Extensive follow-up studies have
yielded accurate measurements of chemical compositions and molecular
gas content for this neutral gas cross-section selected sample from
$z\approx 5$ to $z\approx 0$ (Sect.~\ref{sec:DLAstats}).  Combining
galaxy surveys with absorption-line observations of gas around
galaxies has enabled comprehensive studies of baryon cycles between
star-forming regions and low-density gas over cosmic time.  DLAs,
while being rare as a result of a small cross-section of neutral
medium in the Universe, have offered a unique window into gas dynamics
and chemical enrichment in the outskirts of star-forming disks
(Sect.~\ref{sec:DLAgals}), as well as star formation physics at high
redshifts (Sect.~\ref{sec:sfr}).  Observations of strong
\lya\ absorbers and associated ionic transitions around galaxies have
also demonstrated that galaxy mass is a dominant factor in driving the
extent of chemically enriched halo gas and that chemical enrichment is
well confined within galactic haloes for both low-mass dwarfs and
massive galaxies (Sect.~\ref{sec:cgm}).

With new observations carried out using new, multiplex instruments,
continuing progress is expected in further advancing our understanding
of baryonic cycles in the outskirts of galaxies over the next few
years.  These include, but are not limited to: (1) direct constraints
for the star formation relation in different environments (e.g.,
\citealt{Gnedin:2010}), particularly for star-forming galaxies at $z\gtrsim 
2$ in low surface density regimes of $\Sigma_{\rm SFR}<
0.1\,\msol\,{\rm yr}^{-1}\,{\rm kpc}^{-2}$ and $\Sigma_{\rm
  gas}\approx 10-100\,\msol\,{\rm pc}^{-2}$; (2) an empirical
understanding of galaxy environmental effects in distributing heavy
elements to large distances based on deep galaxy surveys carried out
in a large number of QSO fields (e.g., \citealt{Johnson:2015}); and (3) a
three-dimensional map of gas flows in the circumgalactic space that
combines absorption-line kinematics along multiple sightlines with
optical morphologies of the absorbing galaxies and emission
morphologies of extended gas around the galaxies (e.g.,
\citealt{Rubin:2011, Chen:2014, Zahedy:2016}).  Wide-field
IFUs\index{integral field unit} on existing large ground-based
telescopes substantially increase the efficiency in faint galaxy
surveys (e.g., \citealt{Bacon:2015}) and in revealing extended low
surface brightness emission features around high-redshift galaxies
(e.g., \citealt{Cantalupo:2014, Borisova:2016}).  The {\it James Webb Space
Telescope} ({\it JWST})\index{James Webb Space Telescope}, which is
scheduled to be launched in October 2018, will expand the sensitivity
of detecting faint star-forming galaxies in the early Universe.
Combining deep infrared images from {\it JWST} and CO (or dust continuum)
maps from ALMA\index{ALMA} will lead to critical constraints for the
star formation relation in low surface density regimes.

\begin{acknowledgement} 

The author wishes to dedicate this review to the memory of Arthur
M.\ Wolfe for his pioneering and seminal work on the subject of damped
\lya\ absorbers and for inspiring generations of scientists to pursue
original and fundamental research.  The author thanks Nick Gnedin,
Sean Johnson, Rebecca Pierce, Marc Rafelski, and Fakhri Zahedy for
providing helpful input and comments.  In preparing this review, the
author has made use of NASA's Astrophysics Data System Bibliographic
Services.

\end{acknowledgement}

\bibliography{bibcodes-clean}


\end{document}